\documentclass[twocolumn]{aastex62}
\graphicspath{{./}{figures/}}

\accepted{\today}
\accepted{to be published in ApJ}
\shorttitle{Testing CR Propagation Scenarios}
\shortauthors{Silver and Orlando}
\begin{document}

\title{Testing Cosmic-Ray Propagation Scenarios with AMS-02 and Voyager Data}

\correspondingauthor{Elena Orlando, orlandele@gmail.com}
%\email{orlandele@gmail.com}

\author{Ethan Silver}
\affil{University of California Berkeley, CA-94720, Berkeley, USA}
\affil{Eureka Scientific,
 CA-94602, Oakland, USA}

\author{Elena Orlando}
\affil{University of Trieste, Department of Physics, 34127 Trieste, Italy}
\affil{Istituto Nazionale di Fisica Nucleare - INFN, 34127 Trieste, Italy}
\affil{Stanford University,
CA-94305, Stanford, USA}
\affil{Eureka Scientific,
 CA-94602, Oakland, USA}



\begin{abstract}
AMS-02 on board the ISS provides precise measurements of Cosmic Rays (CR) near Earth, while Voyager measures CR in
the local interstellar medium, beyond the effects of solar modulation. 
Based on these data, we test and revise various CR propagation scenarios under standard assumptions: pure diffusion, diffusion with convection, diffusion with reacceleration, and
diffusion with reacceleration and convection.
We report on the scenarios’ performance against CR measurements, aiming to limit the number of model parameters as much as possible. 
For each scenario we find parameters that are able to reproduce Voyager and AMS-02 data for the entire energy band for all the CR species tested. 
Above several GV we observe a similar injection spectral index for He and C, with He harder than H. Some scenarios previously disfavored are now reconsidered. For example, contrary to usual assumptions, we find that the pure diffusion scenario does
not need an upturn in the diffusion coefficient at low energy, while it needs the same number of low-energy
breaks in the injection spectrum as diffusive-reacceleration scenarios. We show that scenarios differ in modeled spectra
of one order of magnitude for positrons at $\sim$1 GeV and of a factor of 2 for antiprotons at several GV. 
The force-field approximation describes well the AMS-02 and Voyager spectra analyzed,
except antiprotons. We confirm the $\sim$10 GeV excess in the antiproton spectrum for all scenarios. Also, for all scenarios,
the resulting modulation should be stronger for positrons than for nuclei, with reacceleration models requiring  much larger
modulation.
\end{abstract}



\section{Introduction} 
\label{sec:intro}

Cosmic ray (CR) spectra measured by space-based instrumentation is a crucial part of understanding CR propagation throughout the Galaxy. Two valuable sources of CR measurements are the Cosmic Ray Subsystem of Voyager 1 that in the latest 10 years has begun to measure interstellar low-energy CRs \citep{voy1}, and the AMS-02 instrument which has provided high-quality CR data at higher energies since its installation in 2011 \citep{ams02_first}. Among many, additional experiments that add to the available Galactic CR data include the Dark Matter Particle Explorer (DAMPE), the Calorimetric Electron Telescope (CALET), the Cosmic-Ray Isotope Spectrometer on board the Advanced Composition Explorer (ACE-CRIS), and the Payload for Antimatter Matter Exploration and Light-nuclei Astrophysics (PAMELA).
With access to a set of high-quality data, there have been a number of proposed CR propagation models that attempt to predict and explain observed CR spectra (see for example an extensive, though not recent, review in \citep{cr_overview}). There are a wide variety of different propagation models based on different overall propagation scenarios, assumptions about parameters or restrictions placed on them, and datasets used. Research over decades has helped to develop CR propagation models for observed CR spectra, but a number of significant challenges remain \citep{cr_challenges, lipari_article}.
One of the broadest challenges concerns the required complexity of models to explain available CR data. Of course, it is desirable to be able to fit the data with the simplest models, but it is quite common for simple models to be able to fit the data with notable remaining differences with some observations \citep{orlando_cr,schroer_light_model, evoli_model}. On the other hand, other methods use more complex models with many parameters, such as freeing the injection spectrum for all of the primary elements, to achieve the best possible agreement with a wide range of data \citep{boschini2}. Other approaches (\cite{inj_univers,fluka_dragon2}) find that different injection spectra for groups of primaries are in fact required to explain the spectra observed by AMS-02, possibly due to different properties of sources.

The major constraints on propagation scenarios are provided by the two main groups of CR species: the primary species (in particular, H, He, C, O, Ne, Mg, and Si), which are mostly directly produced by CR sources, and the secondary species (including Li, Be, and B), which are in theory totally produced by collisions and the decay of primary CRs \citep{dragon2_seconds}. The primary CR species help to inform details about the generation of CRs. In addition, measurements of secondary species such as Be and B are quite useful to inform details about propagation and to further refine models \citep{evoli_be}. Of particular importance to propagation models is the boron to carbon ratio (B/C), which is commonly used to strongly constrain the diffusion coefficient, so recent precise measurements of the B/C ratio by AMS-02 and other experiments are invaluable to CR modeling \citep{ams02_review}. Recent works have confirmed  that AMS-02 B/C data can be used to fit the diffusion, reacceleration, and convection terms \citep{genolini_bc1, genolini_bc2}, as usually done.
In addition, two other important CR secondaries are positrons and antiprotons \citep{boudaud_antiprotons}. AMS-02 has provided high-quality measurements of positron and antiproton fluxes, yet these are not fully explained by current models of CR transport \citep{cr_overview,  
ams02_review}. For example, work in \citep{lipari_e+_pbar} suggests that both positrons and antiprotons share a common origin as secondaries from hadronic interactions. Also, antiprotons can be valuable to understand because their spectra can be affected by dark matter signatures, providing a potential window into constraining models of dark matter \citep{doetinchem_antiprotons, kahlhoefer_pbar, cuoco_2019}. However, current uncertainties in the solar modulation and CR cross sections may limit the significance of dark matter signals in antiproton data \citep{calore_pbar, cuoco_2019}.

While in the literature there is not agreement on the best propagation parameters, there is a consensus that CR transport can be described by diffusion, reacceleration, convection, or a combination of them. In this work, we test the various standard propagation scenarios with the latest available CR measurements. By revising different propagation scenarios with consistent assumptions, we confirm or disprove common truths that have been accepted so far. Also, we explore the performance of different models, and how the data constrains and informs their parameters, in order to compare which models are favored most strongly by current data.  We are more interested in testing and revising baseline propagation scenarios (diffusion, reacceleration, convection) than in tightly constraining propagation model parameters.

Section \ref{sec:Methods} focuses on the methods, describing the model preparation, the dataset used, and the fitting procedure. Section \ref{sec:Results} discusses the results for each model scenario and the resulting parameters with particular emphasis on the diffusion coefficient and the injection spectral of the various species. It also presents the resulting positrons and antiprotons for the various models and their uncertainties due to the propagation scenario assumed.

\section{Methods}
\label{sec:Methods}

This section describes the propagation code used for the modeling, details on the modeling, the CR dataseet used, and the fitting procedure.

\subsection{ GALPROP Propagation Code}
The GALPROP code\footnote{GALPROP is available from http://galprop.stanford.edu and from https://gitlab.mpcdf.mpg.de/aws/galprop.} is a numerical tool developed over decades \citep[e.g.][]{galprop2, galprop3, cr_overview} to calculate Galactic CR propagation and associated photon emission from gamma rays \citep[e.g.][]{ackermann_halo_size, strong_2010}, through x-rays \citep[e.g.][]{strong_2005, orlando_cr}, to radio and microwaves \citep[e.g.][]{orlando_strong_halo}. Other CR propagation codes are, for example, DRAGON \citep{Evoli_2017}, PICARD \citep{Picard}, USINE \citep{Usine}. %GALPROP numerically solves the CR transport equation, which governs CR propagation, accounting for diffusion, convection caused by galactic wind, reacceleration in the interstellar medium, fragmentation, and other processes.
The GALPROP code computes CR propagation by numerically solving the
CR transport equation  over a grid in coordinates $(R, z, p)$, where $R$ is the radius from
the Galactic centre, $z$ is the height above the Galactic plane, and $p$ is the particle momentum.
The transport equation is described by the following formulation:

\newcommand{\drv}[2]{\frac{\partial #1}{\partial #2}   }
\newcommand{\opdrv}[1]{\frac{\partial}{\partial #1}  }
\newcommand{\ddrvm}[3]{\frac{\partial^{2}  #1}{\partial #2 \partial #3}  }
\newcommand{\ddrv}[2]{\frac{\partial^{2}  #1}{{\partial #2}^2}   }
\newcommand{\Dpp}{D_{pp}}
\newcommand{\Dxx}{D_{xx}}
\newcommand{\ddp}{{\partial\over\partial p}}

\begin{eqnarray}
{\partial \psi (\vec r,p,t) \over \partial t} 
&= &
  q(\vec r, p, t)                                             
   + \vec\nabla \cdot ( D_{xx}\vec\nabla\psi - \vec V\psi )   \nonumber  \\
&   +& \ddp\, p^2 D_{pp} \ddp\, {1\over p^2}\, \psi                  
   - {\partial\over\partial p} \left[\dot{p} \psi
   - {p\over 3} \, (\vec\nabla \cdot \vec V )\psi\right]  + \nonumber  \\
&   - & {1\over\tau_f}\psi - {1\over\tau_r}\psi\ 
\label{eq2}
\end{eqnarray}
where, the terms on the right side represent respectively: CR sources (primaries and secondaries), diffusion, convection (Galactic wind), diffusive reacceleration by CR scattering in the interstellar medium, momentum losses (due to ionization, Coulomb interactions, bremsstrahlung, inverse Compton and synchrotron processes), nuclear fragmentation and radiative decay. $\psi (\vec r,p,t)$ is the CR density per unit of total particle momentum $p$ at position $\vec r$. $\Dxx$ is the spatial diffusion coefficient and is a function of the rigidity R as $Dxx \propto D_0(R/R_0)^\delta$, with $D_0$ normalization at the reference rigidity $R_0$. More specifically, with a break in the diffusion coefficient such that $\delta = \delta_{1,2}$ below/above the reference rigidity, if the reference rigidity is lower than the break rigidity, the diffusion coefficient is described by: $\beta^\eta D_0 (R/R_0)^{\delta_1}$ for $R<R_{break}$ and $\beta^\eta D_0 (R_{break}/R_0)^{\delta_1}(R/R_{break})^{\delta_2}$ for $R>R_{break}$, with $\beta=v/c$. Section \ref{sec:Dxx} contains a deeper discussion on the diffusion coefficient. 
%In our scenarios the diffusion coefficient is isotropic. 
Diffusive reacceleration is described as diffusion in momentum space
and is determined by the coefficient $\Dpp$\ related to  $\Dxx$ by $\Dpp\Dxx\propto p^2$. The convection velocity (in z-direction only) is assumed to increase linearly with distance from the plane (dV/dz $>$ 0 for all z). Because the wind cannot blow in both directions, this formulation requires a zero velocity in z=0, i.e. $V=dV/dz*dz$.

CR propagation models can be evaluated by running GALPROP on their propagation parameters, defined in a configuration file, and GALPROP allows for the exploration of the effects of each parameter on the resulting CR propagation for given gas and CR source distriution.

\subsection{Propagation Models}
The baseline propagation scenarios evaluated and tested in this work are PD (plain diffusion), DRE (diffusion and reacceleration), DRC (diffusion, reacceleration, and convection), and DC (diffusion and convection). DRC models were recently found to be able to agree quite well with AMS-02 spectra \citep[e.g.][]{orlando_cr, boschini2, genolini_bc2}, and successful examples of the simpler DRE and PD models have also been obtained \citep[e.g.][]{orlando_cr, genolini_bc2}.

In order to explain the observed spectral break at a few hundreds of GV, models are split into two physically-based scenarios: the “injection” scenario in which the change in the observed spectral index is caused by a break in the injection CR spectral indices, and the “propagation” scenario in which the change in the observed spectral index is caused by a break in the diffusion coefficient. Compared against current AMS-02 measurements, the propagation scenario has been recently preferred \citep{genolini_break, evoli_model}, so we use the “propagation” scenario for our models. This choice does not affect our results because, as also pointed out in \cite{blasi_2017}, none of the conventional propagation parameters such as low-energy diffusion coefficient, convection velocity, and Alfven velocity are degenerate with this high-energy break.

We test 4 propagation scenarios (DRC, DRE, PD, and DC), for a total of 6 different propagation models. For the DRC scenario, we test three versions: “DRC1”, which only has one low-energy break in the injection spectral index of each species, “DRC2”, which has two low-energy breaks in the injection spectral indices, and “DRC\_conv”, a DRC scenario with a high gradient of the convection velocity $dV_\textrm{conv}/dz$ above 50 km/s/kpc, as recently assumed by magneto-hydrodynamic simulations of galactic outflows, whereas values of around 10-12 km/s/kpc have been more typical in the past.  The other scenarios (DRE, PD, and DC) are all allowed to have two low-energy breaks in the injection spectral indices. 
For all scenarios no high-energy spectral index break is assumed. Even though the PD scenario are usually required to have an additional break in the diffusion coefficient at a few GV with respect to DR and DRC scenarios  \citep[e.g.][]{cr_overview, grenier_strong_2015, strong_2010, cummings_2016}, none of the model scenarios in this work are allowed to have such a break. Low-energy spectral and diffusion breaks are discussed further in Section \ref{sec:low_energy_break}.

The GALPROP runs of each model are performed on a spatial grid with an X and Y range of ±20 kpc and a halo size of Z in the range ±4 kpc with $\Delta$R=1.0 kpc, $\Delta$Z=0.2 kpc, and $\Delta$X=$\Delta$Y=0.5 kpc. We checked that this grid is sufficient to produce accurate results. For example, change to a finer grid of $\Delta$R=0.2 kpc, $\Delta$Z=0.1 kpc, and $\Delta$X=$\Delta$Y=0.2 kpc produces similar spectra, with the greatest deviation being up to 2\% in a few species. The models used are 2D models. We further checked that switching to 3D models, although requiring a large increase in computational cost, also produces changes in the spectra by at most about 2\%. The energy range used is from 1 MeV/nucleon to 100 TeV/nucleon, which results in accurate calculations in the ranges of the tested CR measurements.
With the best galactic halo size still undergoing active research especially because of its degeneracy with the diffusion coefficient, we choose a standard fixed halo size of $z=4$~kpc for all models. While propagation parameter values may be affected by the choice of the halo size, our conclusions are not. Recent research that used $^{10}$Be/$^9$Be data obtain $z=3.8^{+2.8}_{-1.6}$ kpc in \citep{be_halo} or $z=5^{+3}_{-2}$ in \citep{weinrich_halo}, and other work finds a lower bound of $z\geq 3$ \citep{be_halo_bound}, or simply supports a lower halo size of a few kpc \citep{evoli_halo}. 
On the other hand, some researches supports a larger halo size ($z\geq 4$ kpc up to $z=10$ kpc) \citep{dragon2_seconds, ackermann_halo_size, orlando_strong_halo, evoli_be}.

For our models we assume the magnetic field configurations as optimized in \citet{orlando_cr,orlando_prd},
the standard CO and HI gas density models\footnote{The files $rbands\_co10mm\_v3\_2001\_hdeg.fits.gz$ for the CO distribution and $rbands\_hi12\_v5\_hdeg\_zmax1\_Ts125.fits.gz$ for the HI distribution are used.} as released with the GALPROP code, and the CR source distribution\footnote{The source distribution is assumed  as a function of Galactocentric radius $r$ and the vertical distance $z$ to the Galactic plane it follows the formula: $(r/8.5)^{0.475} * exp(-2.1657*(r  -8.5)/8.5) * exp(-|z|)/0.2)$, and it is constant for r~$\geq$~10kpc.} as in \citet{strong_2010} and  also used in \citet{orlando_strong_halo}, which is based on \cite{Lorimer2006} and it is consistent with gamma-ray data and the previously cited works on the magnetic fields. Other source distributions  could also lead to a good fit to Voyager data \cite[e.g][]{Schlickeiser}.

\begin{table*}%[!hp]
\centering
\caption{CR Datasets used}
%\begin{ruledtabular}
\label{table:datasets}
\begin{tabular}{ccccccc}
\hline
Species & Instrument & Collection Period\\
\hline
H,He,B,C,O,Ne,Mg,Si & AMS-02 & 2011/05-2018/05\\
B,C,O,Ne,Mg,Si & Voy1-HET-Bend & 2012/12-2014/12\\
H,He & Voy1-HET & 2012/12-2015/06\\
H,He,B,C,O,Ne,Mg,Si & Voy1-HET-Aend & 2012/12-2015/06\\
H,He,B,C,O,Ne,Mg,Si & Voy1-LET & 2012/12-2015/06\\
H & DAMPE & 2016/01-2018/06\\
H & CALET & 2015/10-2018/08\\
B/C & ACE-CRIS & 1997/08-2010/01\\
\hline
\end{tabular}
%\end{ruledtabular}
\end{table*}

\subsection{CR Datasets}
Our main dataset of CR are low-energy measurements outside the solar system from Voyager 1 (H, He, B, C, B/C, O, Ne, Mg, and Si) \citep{cummings_2016}, and precise high-energy measurements from AMS-02 (H, He, B, C, B/C, O, Ne, Mg, and Si) \citep{ams02_review}. Additional datasets used are H measurements from DAMPE \citep{dampe} and CALET \citep{calet}, and B/C measurements from ACE-CRIS \citep{ace_cris}. The CR data used is shown in Table \ref{table:datasets}. Most data is taken from \citep{maurin2023}.

\subsection{Parameter Fitting Procedure}
\label{sec:parameter_fitting}
Minuit2 is an optimization library maintained by CERN, written in C++ and originally developed in FORTRAN \citep{minuit}. We use Minuit2’s Migrad algorithm to minimize the error between the output of our  model and the data of CR spectra and their abundances in order to optimize the propagation parameters. Each data point is weighted by the reported error, as well as other considerations for the data that the models should prioritize, such as the B/C spectra, which is desirable for a very close match to the AMS-02 data in order to constrain propagation parameters.
In more detail, following the usual commonly accepted procedure, we assume B to be totally of secondary origin. For each scenario we primarily use B, C, and B/C at first to optimize the overall parameters such as those governing diffusion, reacceleration, and convection, as well as the C injection parameters. Then, the injection spectrum and abundance of each of the primary species (H, He, C, O, Ne, Mg, and Si) are optimized by comparing against the CR spectra of each species. Voyager data is in units of $E_{kin}/nuc$, while most AMS-02 data is in units of rigidity with conversions reported in the papers. Hence, models are fitted to AMS-02 data in rigidity and to Voyager data in $E_{kin}/nuc$. We decide to visualize most of our results in $E_{kin}/nuc$ and where necessary we convert AMS-02 data for illustration. Each model also has the solar modulation fit, although the data unaffected by solar modulation — namely, Voyager data and AMS-02 data at high energies — is weighted more than the low-energy data strongly affected by modulation. To apply solar modulation, we use the common force-field approximation \citep{gleeson_axford_mod}. In detail, chi-squared loss is used, with the effective error on AMS-02 B/C data %reduced by 5X, 
and 10\% error added to AMS-02 data below 45 GV, accounting for the uncertainties due to the solar modulation force-field approximation.
The choice of 45 GV is based on AMS-02 proton and heavier nuclei data that do not show any variation due to modulation above that rigidity.

The propagation parameters, the injection spectra, and the normalization of the species have been left free to vary as further discussed in Section 3. In particular, for each species, the three injection spectral indices (or two for DRC1) are left free for the primary species H, He, C, O, Ne, Mg, and Si. 
Among the propagation parameters are the diffusion coefficient and its slopes for all models, and the Alfven and convection velocity for the models that include them. The energy of the break of the diffusion coefficient, which has been initially included as a free parameter, later in the optimization it has been fixed, being not relevant in this study and not affecting our results and conclusions, as previously discussed. In each scenario propagation parameters are the same for all CR species, which has a stronger physical basis than just assuming ad hoc propagation parameters of injection spectra for each species.
We also include a factor in the diffusion coefficient of $\beta^\eta=(v/c)^\eta$ with a free parameter $\eta$, which improves agreement at low energies for certain models, as used in recent works \citep[e.g.][]{boschini2, genolini_bc1, genolini_bc2,maurin_2010, Evoli_2015, Evoli_2017}. A deeper discussion of this parameter is reported in Section 3.6.

\begin{table*}
%\centering
\caption{GALPROP parameters for all models.}
%\begin{threeparttable}[b]
\begin{ruledtabular}
\begin{tabular}{cccccccc}
%\hline
\hline
Parameter&DRC1&DRC2&DRC\_conv&DRE&PD&DC\\
\hline
\tablenotemark{a}$D_{0}\ (10^{28} cm^{2}s^{-1})$& 4.3118 & 4.4452 & 2.7425 & 4.7776 & 4.5767 & 3.6183\\
\tablenotemark{b}$\delta_{1}$& 0.4124 & 0.4163 & 0.4476 & 0.4052 & 0.4047 & 0.4448\\
\tablenotemark{b}$\delta_{2}$& 0.2117 & 0.2404 & 0.2726 & 0.2315 & 0.1928 & 0.1975\\
$R_{break}$ (GV) & 295.97 & 308.04 & 332.84 & 308.04 & 290.67 & 283.29\\
\tablenotemark{c}$\eta$ & 0.6414 & 0.4373 & 0.7152 & 0.3851 & 0.0004 & 0.8196\\
$V_{Alf}\ (km\ s^{-1})$ & 30.242 & 32.187 & 51.605 & 26.727 & ˙˙˙ & ˙˙˙\\
$dV/dz (km\ s^{-1}kpc^{-1})$ & 10.233 & 6.3482 & 54.047 & ˙˙˙ & ˙˙˙ & 10.022\tablenotemark{d}\\
$\gamma_{0, C}$ & 0.8172 & 0.7286 & 1.9321 & 0.9210 & 0.0862 & -0.5963\\
$\gamma_{1, C}$ & 2.3570 & 2.2043 & 0.5733 & 2.2947 & 1.9944 & 2.0575\\
$\gamma_{2, C}$ & ˙˙˙ & 2.3640 & 2.3726 & 2.3601 & 2.3159 & 2.3114\\
$\rho_{1, C}\ (GV)$  & 1.5711 & 1.2670 & 4.7874 & 1.5347 & 0.6510 & 0.7037\\
$\rho_{2, C}\ (GV)$  & ˙˙˙ & 7.9872 & 6.0414 & 12.141 & 8.0493 & 8.7891\\
$q_{0,^{12}C}/q_{0,H}*10^{6}$ & 3205.2 & 3215.9 & 3040.6 & 3241.9 & 3177.3 & 3279.5\\
$\Phi/\textrm{Modulation}\ (MV)$ & 633 & 622 & 600 & 612 & 368 & 375\\
Unmodulated $\chi^2$ & 581 & 428 & 767 & 430 & 571 & 741\\
%\hline
\label{table:params}
\end{tabular}
\tablenotetext{a}{Normalized at $4 GV$}
\tablenotetext{b}{$\delta_{1}$ below and $\delta_{2}$ above the break rigidity}
\tablenotetext{c}{Factor of $\beta^\eta$ on the diffusion coefficient, where $\beta=v/c$}
\tablenotetext{d}{If the $dV_\textrm{conv}/dz$ is allowed to assume any value, the result is the PD scenario where $dV_\textrm{conv}/dz$ is fit to 0 km/s/kpc}
\end{ruledtabular}
%\end{threeparttable}
\end{table*}

\section{Results}
\label{sec:Results}
\subsection{Best-Fit GALPROP Models}
The best-fit propagation parameters generated by the minimization procedure for each scenario, as well as chi-squared values, are presented in Table \ref{table:params}. Due to the construction the absolute value of the chi-squared can be used to compare models. Chi-squared values are reported by using the data unaffected by solar modulation (unmodulated chi-squared), i.e. Voyager data and AMS-02 data above 20 GV\footnote{Taking 45 GV instead of 20 GV does not change the order of the chi-square of the models} (with 10\% error below 45 GV), except for B/C, which is above 4 GV. The parameters governing the C injection spectrum are also included in the Table. %, due to the importance of matching B, C, and B/C measurements.

\begin{figure*}
    \centering
    \includegraphics[width=.9\textwidth]{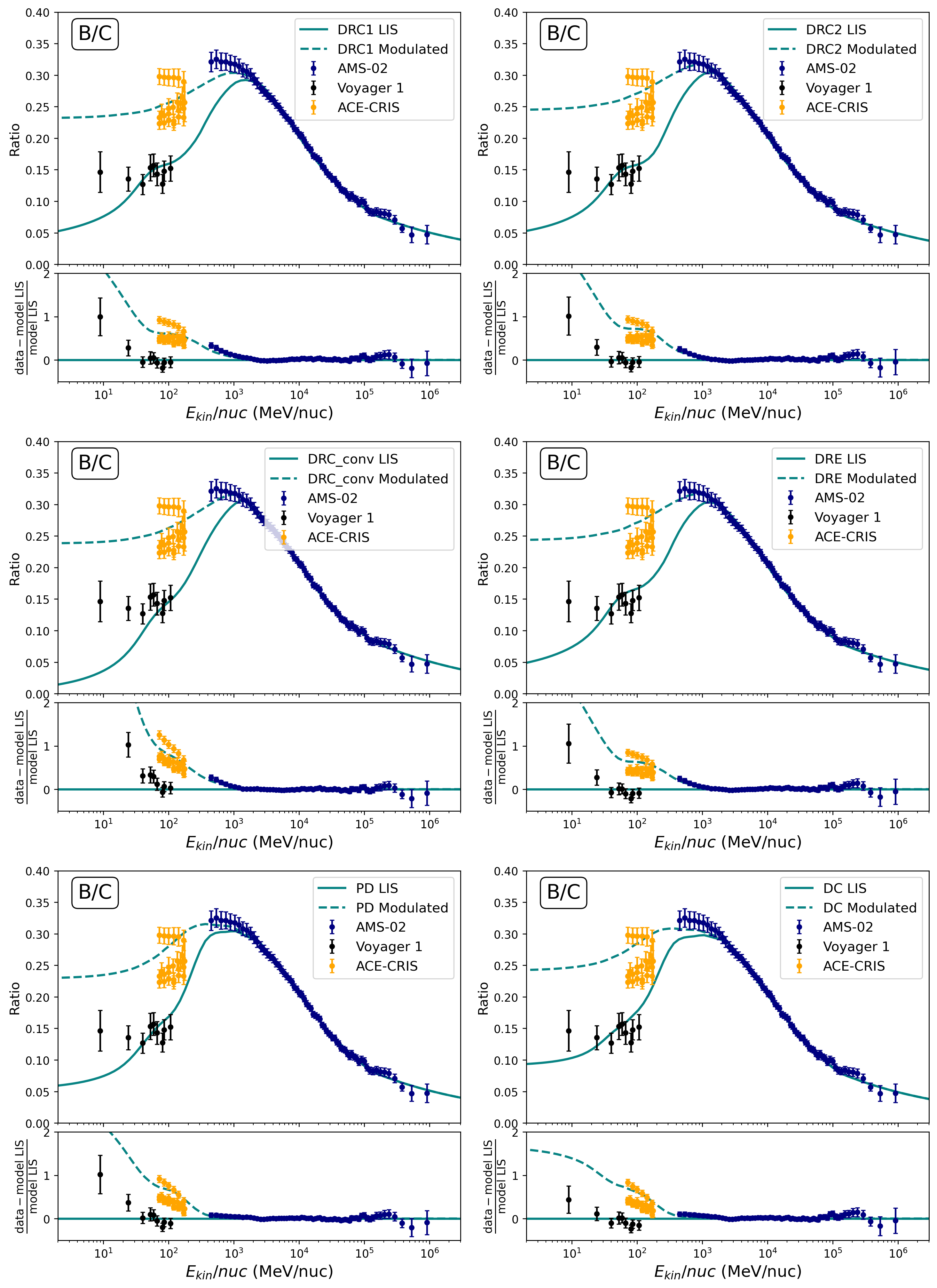}
    \caption{Plots of the B/C ratio in units of kinetic energy for each of the six tested scenarios, top to bottom, left to rigth: DRC1, DRC2, DRC\_conv, DRE, PD, DC. Residuals are also shown. Solid lines represent the LIS, while dashed lines represent the modulated spectra. Data are: AMS-02 data (blue points) \citep{ams02_bc_old}, ACE-CRIS (orange points) \citep{ace_cris},  and Voyager 1 (black points) \citep{cummings_2016}.}
    \label{fig:comb_bc}
\end{figure*}

\begin{figure*}
    \centering
    \includegraphics[width=.9\textwidth]{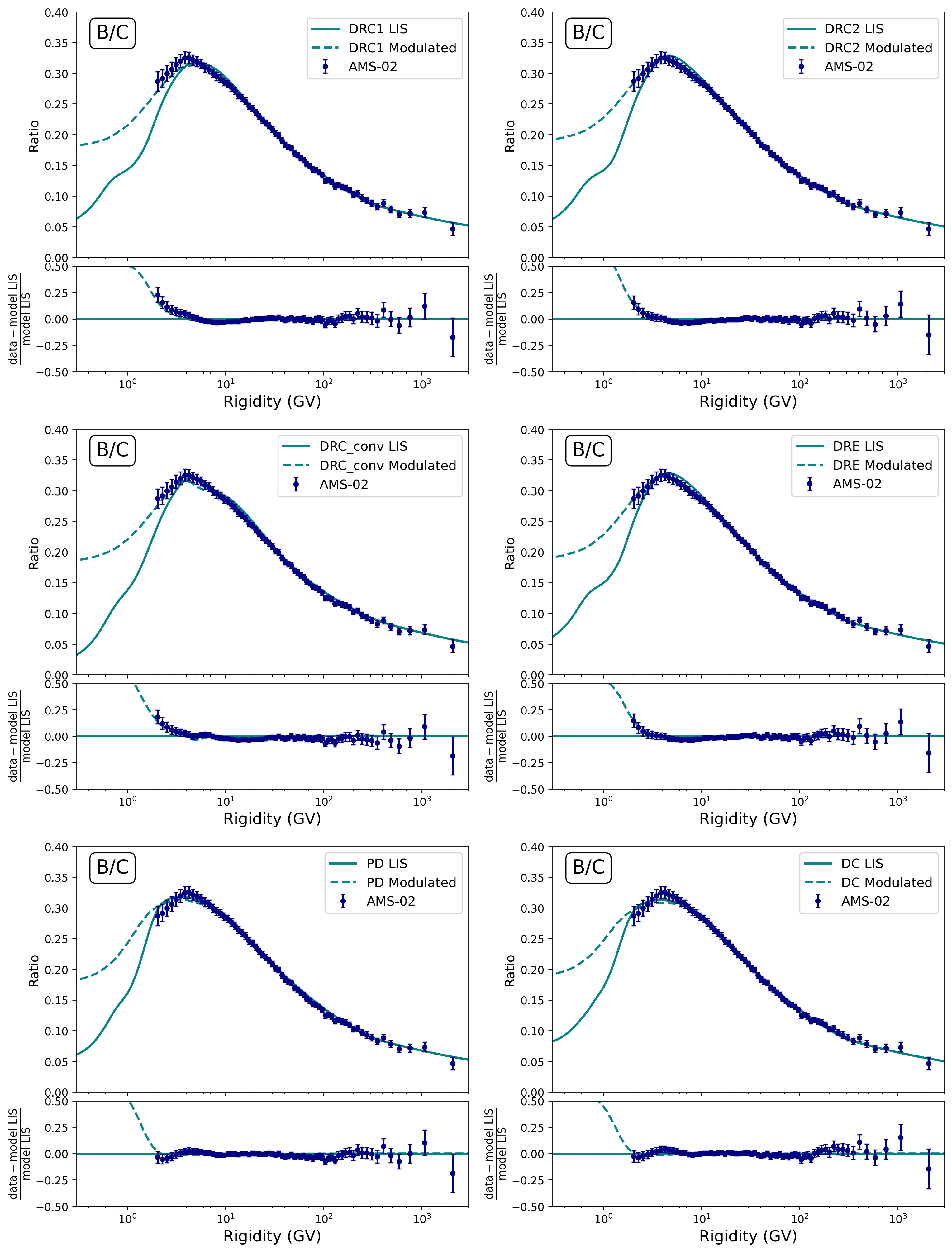}
    \caption{Plots of the B/C ratio in units of rigidity for each of the six tested scenarios, top to bottom, left to rigth: DRC1, DRC2, DRC\_conv, DRE, PD, DC. Residuals are also shown. Shown are the most recent AMS-02 data \citep{ams02_review}. Solid lines represent the LIS, while dashed lines represent the modulated spectra.}
    \label{fig:comb_bcr}
\end{figure*}

Figure \ref{fig:comb_bc} shows the B/C ratio for each model in units of $E_{kin}/nuc$, and Figure \ref{fig:comb_bcr} shows the B/C ratio for each model in units of rigidity. As discussed in Section \ref{sec:parameter_fitting}, all of the AMS-02 data used for fitting, including B/C, is in units of rigidity. However, the Voyager 1 and ACE-CRIS data is in units of $E_{kin}/nuc$. To display them on the same B/C plot, 
we use the AMS-02 B/C data given in \citet{ams02_bc_old}, which provides the spectrum in both units.
Solid lines represent the Local Interstellar Spectra (LIS), while dashed lines represent the modulated spectra.
For all scenario we find parameters that are able to reproduce Voyager and AMS02 data within error bars for H, He, B, C, B/C, O, Ne, Mg, and Si.
The models can all match the AMS-02 data within the error for nearly all data, and at low energies, the modulated spectrum falls within the range of values for ACE-CRIS data, and the LIS spectrum falls within the range of values for Voyager 1 data. All scenarios underpredict the Voyager B/C data point at the lowest energy ($\sim$10 MeV/nucl.). In general, the detailed performance of each model varies in certain areas of the spectrum, in particular at the peak of B/C at $\sim1$ GeV/nuc and the low-energy range. In the following sections we discuss each model in detail.

\subsection{DRC Scenarios}
\begin{figure*}
    \centering
    \includegraphics[width=.665\textwidth]{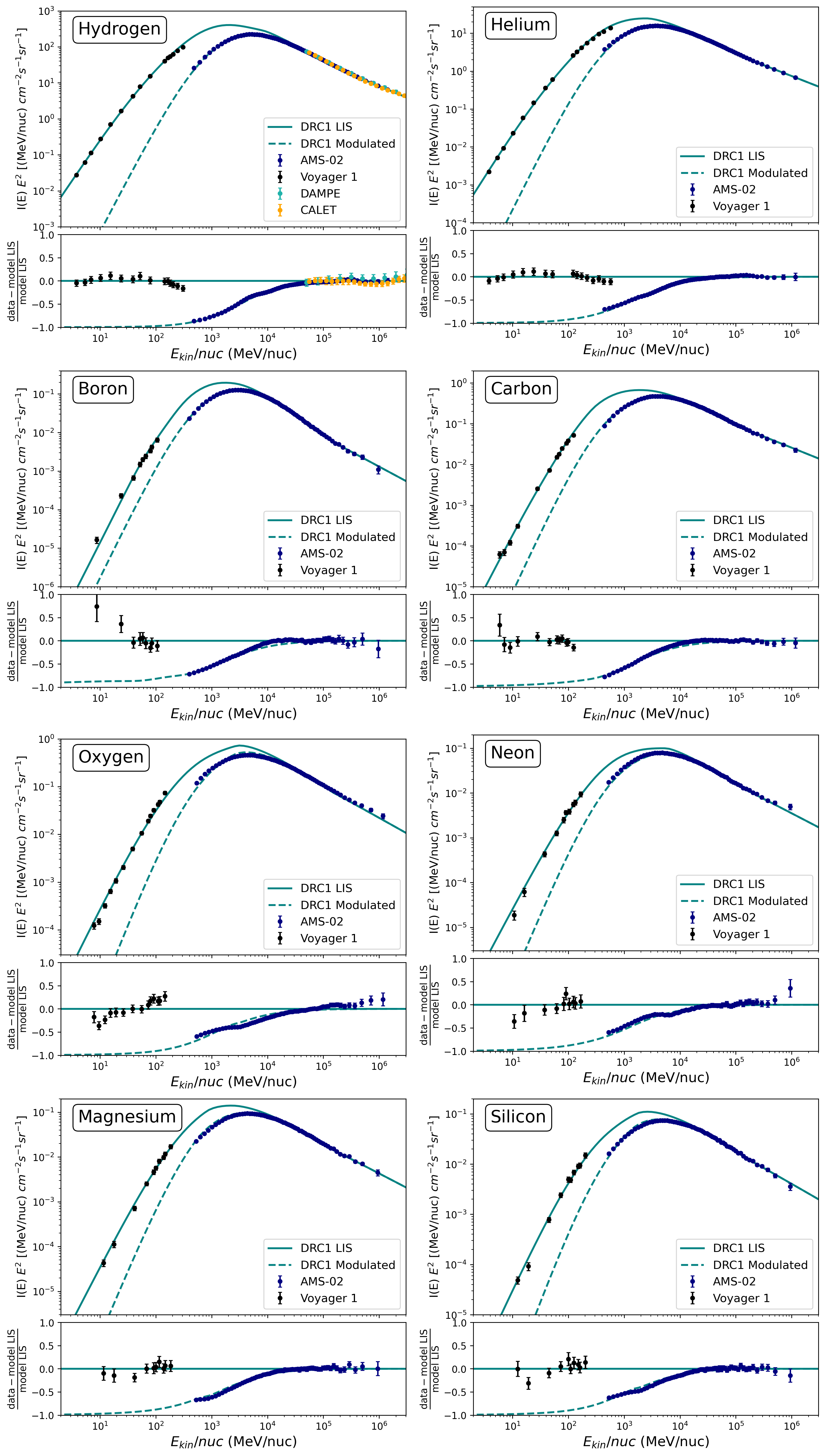}
    \caption{Plots of the CR spectra for each species analyzed for the DRC1 model. Residuals are also shown. Data are: AMS-02 data (blue points) \citep{ams02_bc_old}, Voyager 1 (black points) \citep{cummings_2016}, DAMPE (cyan points) \citep{dampe}, and CALET (orange points) \citep{calet}. Solid lines represent the LIS, while dashed lines represent the modulated spectra.}
    \label{fig:comb_drc1}
\end{figure*}

\begin{figure*}
    \centering
    \includegraphics[width=.665\textwidth]{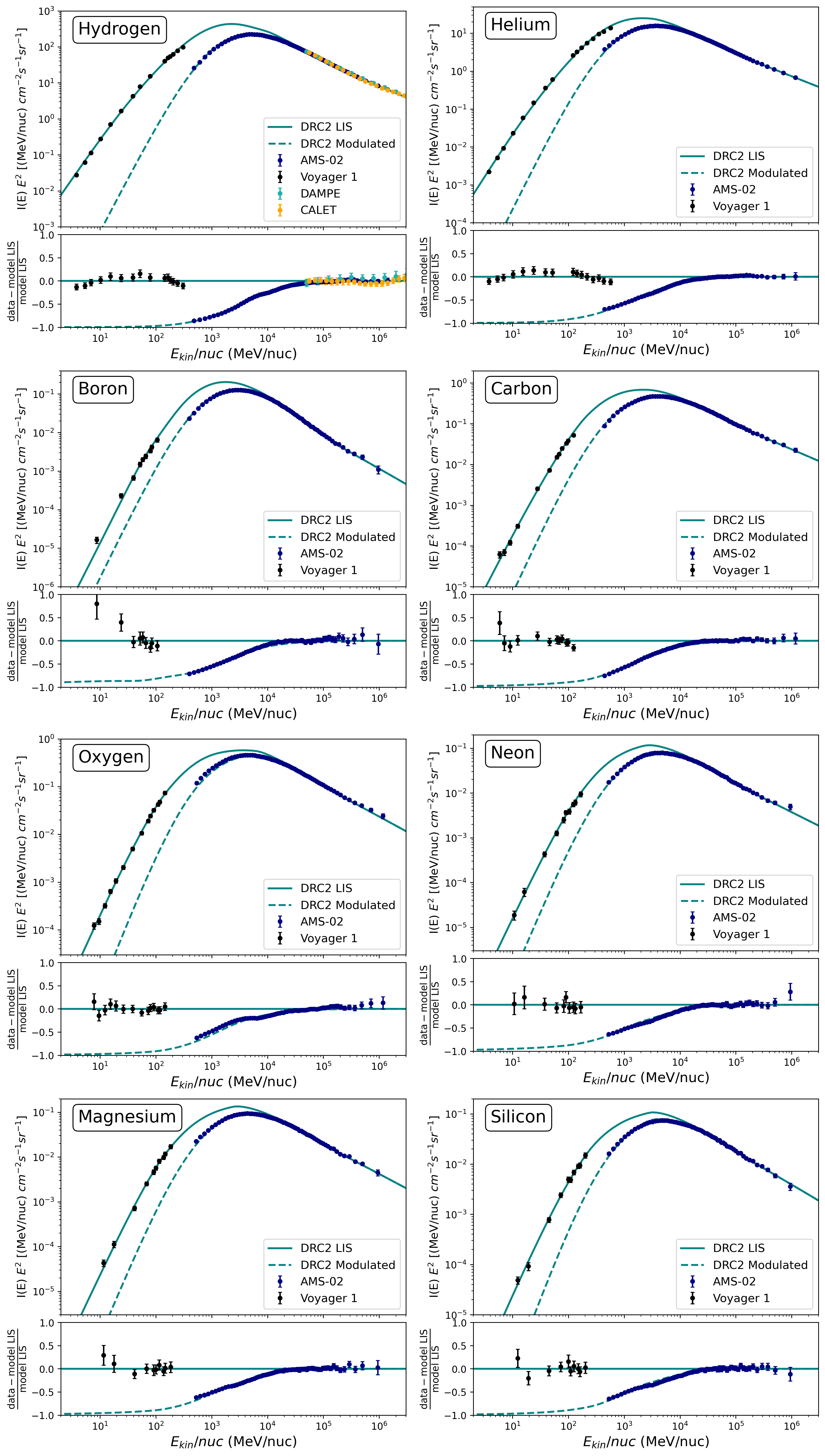}
    \caption{Plots of the CR spectra for each species analyzed for the DRC2 model. Residuals are also shown. Data are: AMS-02 data (blue points) \citep{ams02_bc_old}, Voyager 1 (black points) \citep{cummings_2016}, DAMPE (cyan points) \citep{dampe}, and CALET (orange points) \citep{calet}. Solid lines represent the LIS, while dashed lines represent the modulated spectra.}
    \label{fig:comb_drc2}
\end{figure*}

\begin{figure*}
    \centering
    \includegraphics[width=.665\textwidth]{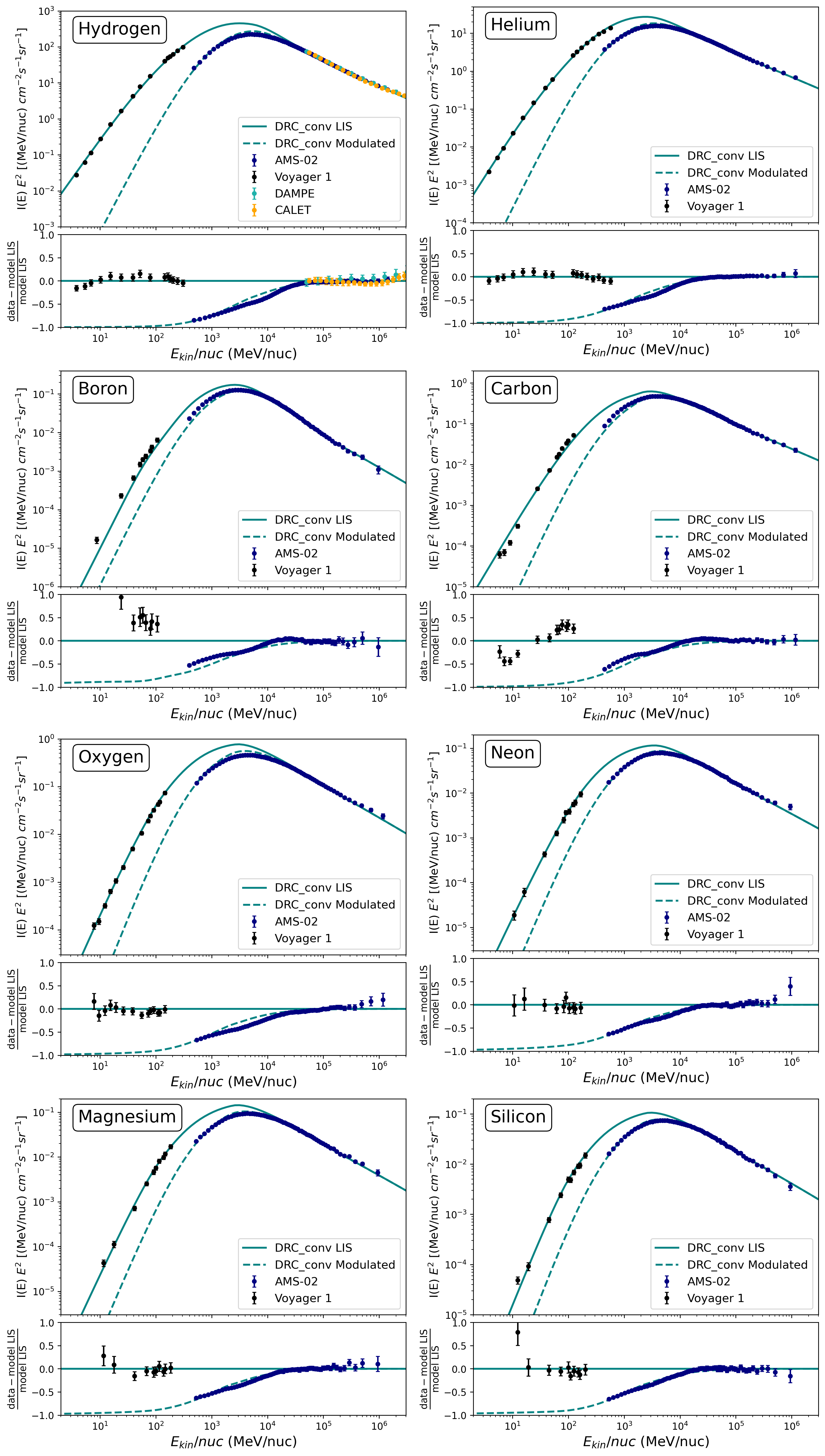}
    \caption{Plots of the CR spectra for each species analyzed for the DRC\_conv model. Residuals are also shown. Data are: AMS-02 data (blue points) \citep{ams02_bc_old}, Voyager 1 (black points) \citep{cummings_2016}, DAMPE (cyan points) \citep{dampe}, and CALET (orange points) \citep{calet}. Solid lines represent the LIS, while dashed lines represent the modulated spectra.}
    \label{fig:comb_drc_conv}
\end{figure*}

DRC scenarios of CR propagation provide the greatest flexibility in propagation modeling by including parameters for diffusion, reacceleration, and convection. Three variants of DRC model are included in our analysis: DRC1, DRC2, and DRC\_conv. The spectra for each species are shown in Figures \ref{fig:comb_drc1}, \ref{fig:comb_drc2}, and \ref{fig:comb_drc_conv}, respectively.
Comparing DRC1 and DRC2, most of the parameters are fit to be similar, with the reacceleration and convection, controlled by parameters $V_{Alfven}$ and $dV_\textrm{conv}/dz$, with values falling in the ranges $\sim$~30~--~32~km~s$^{-1}$ and $\sim$~6~--~10~km~s$^{-1}$kpc$^{-1}$, respectively. The primary difference between the models is in the C injection spectrum. DRC1 only has one low-energy break in the spectral index of C, at $\sim$1.5 GV. DRC2 has two spectral index breaks, which are fit to $\sim$1.3 GV and $\sim$8 GV. Thus, DRC1 and DRC2 differ by 2 model parameters per species (14 total for seven primary species). This addition improves the detail of the low-energy injection spectrum, but makes the model more complicated. This primarily results in DRC2 having a higher B/C spectrum at low-energies, which brings the model into closer alignment with AMS-02 and Voyager 1 measurements. 
DRC\_conv has a similar setup to DRC2, except that $dV_\textrm{conv}/dz$ was limited to be greater than 50 km/s/kpc, in order to investigate the feasibility of such a scenario, which would possibly account for evidence for a Galactic wind \citep[e.g.][]{Everett_2010,pfrommer_2016,hanasz_2009,recchia_2017,Zweibel2013}. This model has stronger convection, as well as reacceleration, with $V_{Alfven}>$50km/s, while the diffusion coefficient is significantly lower than the other DRC models. The resulting C injection spectrum is also atypical at low energies, with an ad hoc hardening 
between $\sim$5 GV and $\sim$6 GV, and then softening again to typical high energy values, which currently would not have a physical explanation (this does not occur for the other models). DRC\_conv has a B/C spectrum comparable to the other DRC models, and generally agrees with AMS-02 and Voyager 1 spectra, although not quite as well. However, DRC\_conv has noticeably worse performance on low-energy B and C individually (Figure \ref{fig:comb_drc_conv}).
Out of the DRC scenarios, DRC2 has the lowest chi-squared, and DRC1 the second least, while DRC\_conv is statistically disfavoured. DRC2 has a better unmodulated chi-squared than DRE, while DRC1 does not, but both have good low-energy performance for B/C (Figure \ref{fig:comb_bcr}). However, this low-energy region has the most uncertainty due to the effects of solar modulation.

\subsection{DRE Scenario} 
\begin{figure*}
    \centering
    \includegraphics[width=.665\textwidth]{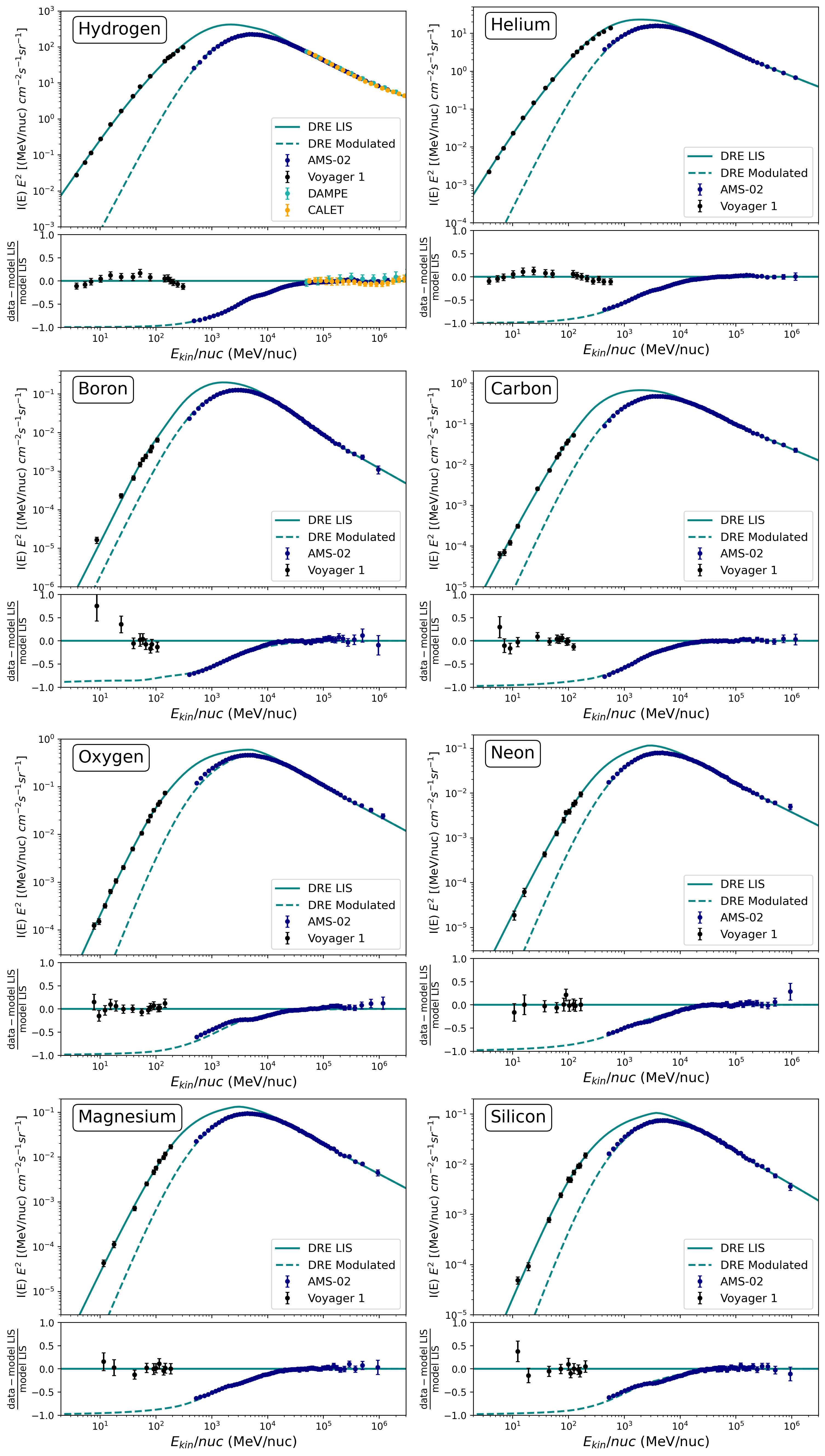}
    \caption{Plots of the CR spectra for each species analyzed for the DRE model. Residuals are also shown. Data are: AMS-02 data (blue points) \citep{ams02_bc_old}, Voyager 1 (black points) \citep{cummings_2016}, DAMPE (cyan points) \citep{dampe}, and CALET (orange points) \citep{calet}. Solid lines represent the LIS, while dashed lines represent the modulated spectra.}
    \label{fig:comb_dre}
\end{figure*}

The DRE scenario has the same setup to DRC2, with two breaks in the spectral indices, but with no convection parameter, just including diffusion and reacceleration. Hence, it has one model propagation parameter less than the DRC2 model and one more than DRC1 (excluding the additional 14 free parameters in the injection spectra).
The spectra for the DRE scenario for each species are shown in Figure \ref{fig:comb_dre}. In comparison to DRC2, DRE has similar reacceleration, with $V_{Alfven}=$ 25 km/s, 
with the diffusion parameters also similar. The low-energy C spectrum is somewhat different, where DRE matches the AMS-02 peak in B/C well. The DRE model has a greater total chi-squared than DRC2, with the second-lowest value among all scenarios. This indicates that DRE scenarios can perform nearly as well on unmodulated data as a DRC2 scenario, which has one additional model parameter.

\subsection{PD scenario} 
\begin{figure*}
    \centering
    \includegraphics[width=.665\textwidth]{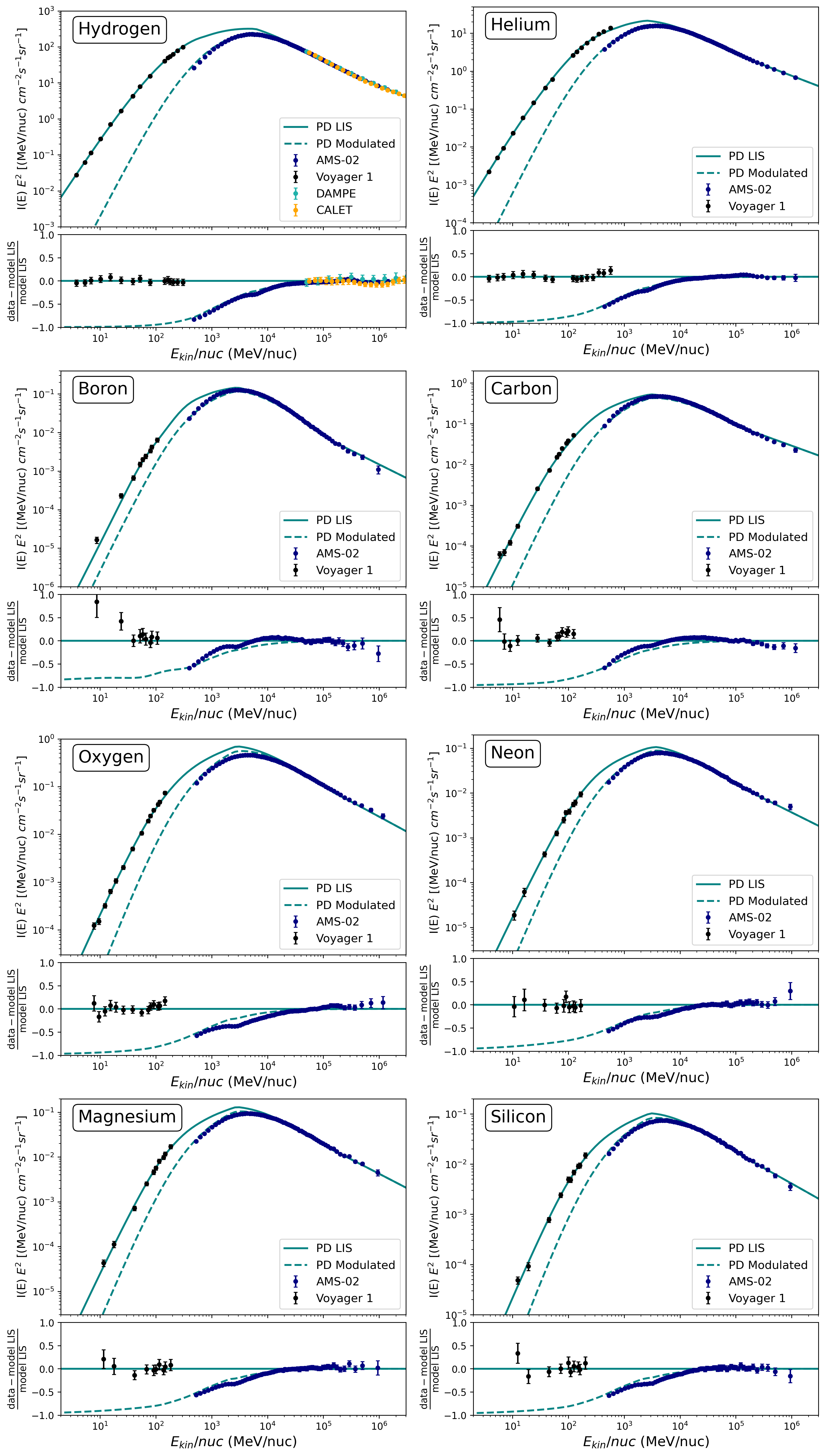}
    \caption{Plots of the CR spectra for each species analyzed for the PD model. Residuals are also shown. Data are: AMS-02 data (blue points) \citep{ams02_bc_old}, Voyager 1 (black points) \citep{cummings_2016}, DAMPE (cyan points) \citep{dampe}, and CALET (orange points) \citep{calet}. Solid lines represent the LIS, while dashed lines represent the modulated spectra.}
    \label{fig:comb_pd}
\end{figure*}

The PD scenario has the same setup to DRC2 and DRE, but does not include convection or reacceleration. It has 2 and 1 model parameters less than the DRC2 and DRE scenarios, respectively. 
The spectra for the PD model are shown in Figure \ref{fig:comb_pd}. The PD model has a diffusion coefficient similar to DRC2 and similar C injection spectrum with slightly harder spectral indices. The model has a parameter of $\eta$ optimized to nearly 0, unlike most of the other models, which helps its low-energy performance for most species. There are some aspects in which the modulated spectrum of PD is not quite as good as DRC2 and DRE, often in the 1 GeV/nuc to 10 GeV/nuc range, where the PD model over-produces some species, especially O. However, this is the energy range where uncertainty due to the solar modulation force-field approximation is large. The peak of B/C is in the 200-300 MeV/nuc range, lower than in other models, where it is around 1 GeV/nuc.  Overall, the PD scenario can still sufficiently model most species’ CR spectra, although with less accuracy than the more complex DRC2 model, but still comparable to DRC1 model.

\subsection{DC scenario} 
\begin{figure*}
    \centering
    \includegraphics[width=.665\textwidth]{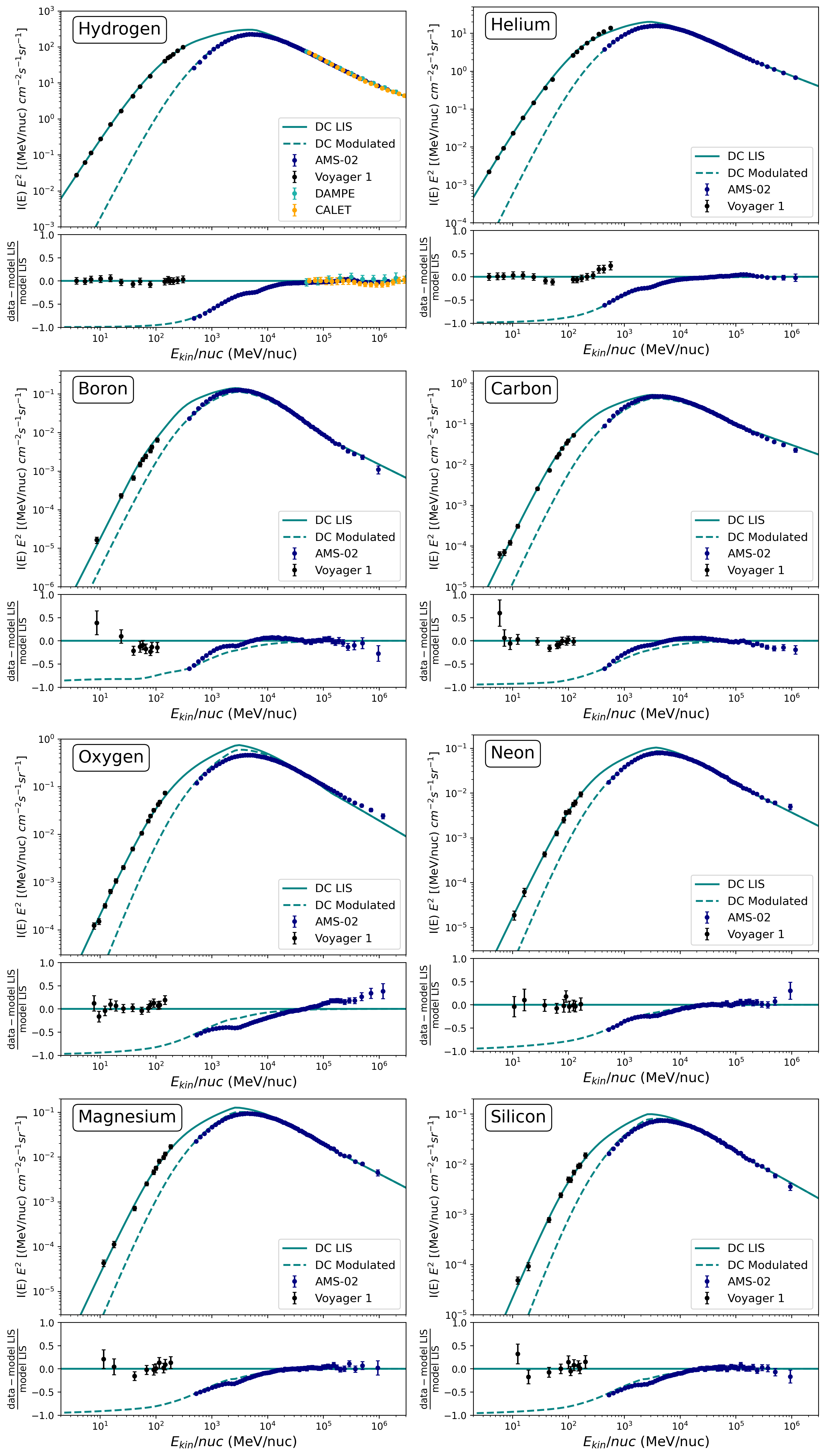}
    \caption{Plots of the CR spectra for each species analyzed for the DC model. Residuals are also shown. Data are: AMS-02 data (blue points) \citep{ams02_bc_old}, Voyager 1 (black points) \citep{cummings_2016}, DAMPE (cyan points) \citep{dampe}, and CALET (orange points) \citep{calet}. Solid lines represent the LIS, while dashed lines represent the modulated spectra.}
    \label{fig:comb_dc}
\end{figure*}

The DC scenario also has the same basic setup as DRC2, DRE, and PD, including convection but not reacceleration. It has the same number of parameters as DRE, and one more than PD. The spectra for the DC scenario are shown in Figure \ref{fig:comb_dc}. However, if the $dV_\textrm{conv}/dz$ is allowed to assume any value, the result is the PD scenario where $dV_\textrm{conv}/dz$ is fit to 0 km/s/kpc. In order to test the sensitivity to the initial values of the fit, the optimization was started with different values, such as a lower value of $D_0$, and a higher value of $\delta_1$. The minimum of error found, with $dV_\textrm{conv}/dz$ at about 10 km/s/kpc, is only a local minimum, as the value of chi-squared is strictly higher for the DC model than for PD, and this characteristic can also be observed in the predicted CR spectra. B/C is overall similar to that of PD, but DC performs a little worse at the peak, and is also worse at low-energies. Other weaknesses of the PD model, such as parts of the O spectrum, are replicated for DC, which is less ideal than other models on most species.  Overall, the DC scenario is disfavored over the PD scenario.

\subsection{The Diffusion Coefficient}
\label{sec:Dxx}
The spatial diffusion coefficient is typically assumed to be a power law in particle rigidity. It increases with rigidity, which is responsible for the observed CR spectra being steeper than the injection spectra. The spatial diffusion coefficient is often written as $D_{xx}$ to distinguish it from $D_{pp}$, the diffusion coefficient in momentum. Standard theoretical models of CR diffusion generally give the rigidity dependence $D_{xx}$ $\propto$ $R^\delta$ \citep{cr_overview}
with $\delta=1/3$ for a Kolmogorov turbulence spectrum \citep{kolmogorov_1941} and $\delta=1/2$ for a Kraichnan turbulence spectrum \citep{iroshnikov_1964,kraichnan_1965}.  To explain the recently observed CR hardening at a few hundred GV, models based on propagation scenarios require a break in the diffusion coefficient with a lower value of $\delta$ above the rigidity break.
Our models use a homogeneous and isotropic diffusion coefficient throughout the Galaxy with the following form as a function of the rigidity $R$: $D_{xx}=D_0~\beta^\eta ~R^\delta$, where $D_0$ is the value of the diffusion coefficient at a given rigidity, $\delta$ is the slope, and $\beta = (v/c)$ affects the low-rigidity regime only. 
As in recent works \citep[e.g.][]{boschini2, genolini_bc1, genolini_bc2,maurin_2010, Evoli_2015, Evoli_2017},
the factor $\eta$ has the effect of increasing the diffusion coefficient at low energies for values of $\eta<1$, while for $\eta=1$, $\beta^\eta = (v/c)$, as traditionally assumed. This parameter has no effect above a few GV in rigidity. A negative $\eta$ allows accounting for physical non linear phenomena at low energies, such as the turbulent dissipation and wave damping that produce a very sharp rise of the diffusion coefficient at rigidity less than around 1.5 GV \citep{ptuskin_damping, cr_overview}.

\begin{figure*}
    \centering
    \includegraphics[width=.7\textwidth]{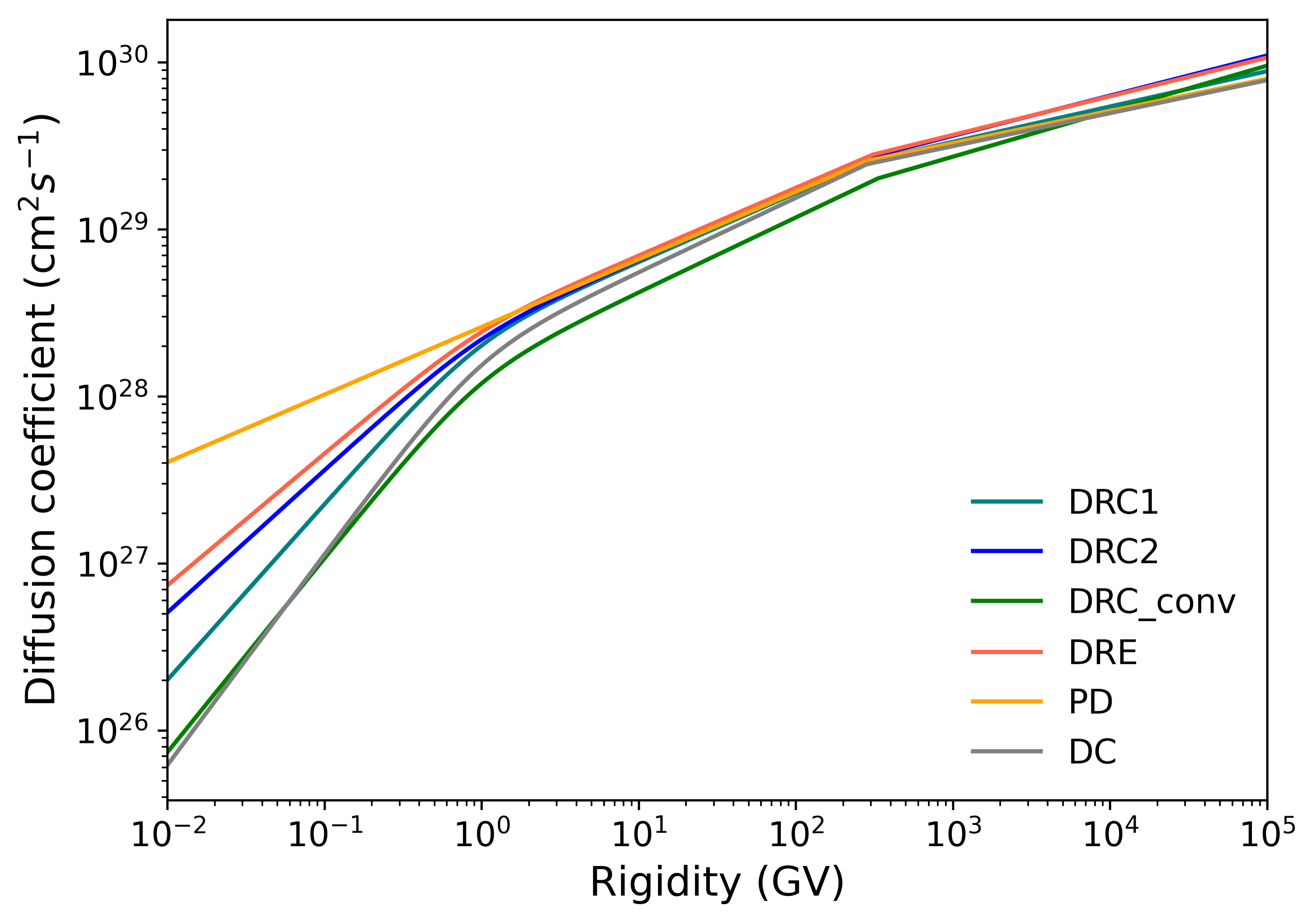}
    \caption{The diffusion coefficient as a function of rigidity for A/Z=1.}
    \label{fig:dxx}
\end{figure*}

\begin{figure*}
    \centering
    \includegraphics[width=.7\textwidth]{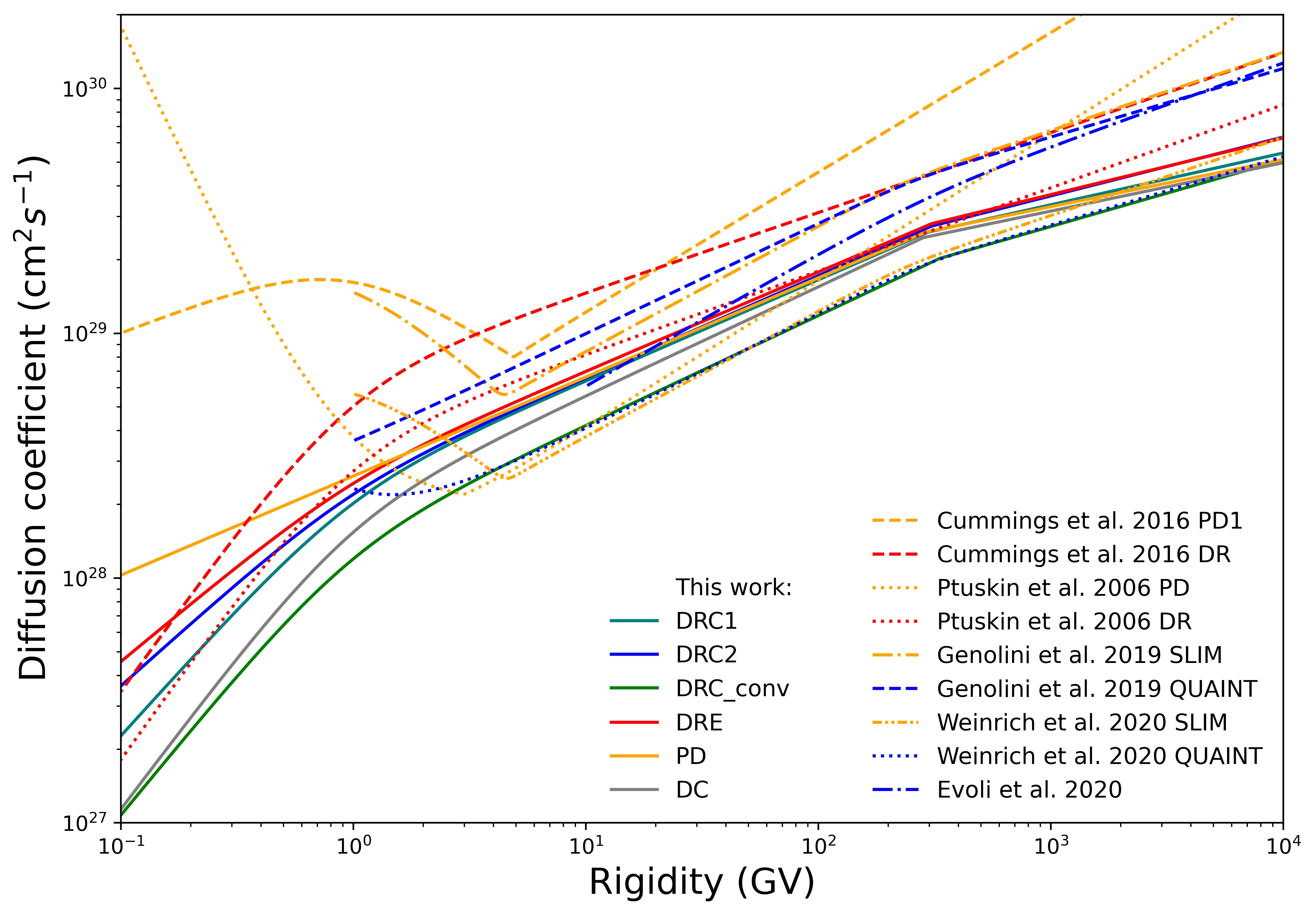}
    \caption{The diffusion coefficient as a function of rigidity for A/Z=1 for the present work together with previous works as listed in the legend.}
    \label{fig:dxxAll}
\end{figure*}

Table \ref{table:params} reports the values of the diffusion coefficient as obtained for our best-fit scenarios.
The models DRC1, DRC2, DRE, and PD all give a value of $D_0$  of about $4.3-4.8 \times 10^{28} cm^2 s^{-1}$ at 4 GV, while the models DRC\_conv and DC have lower values of $D_0$. All our models give $\delta$ in the range of 0.4 to 0.45 below the high-energy break, which matches typical values for a range of CR propagation models, between the theoretical Kolmogorov and Kraichnan scenarios. 
These values for $\delta$ are all above the value of $1/3$ in the theoretical Kolmogorov reacceleration model as preferred in the past \citep{cr_overview},
but in line with more recent studies with AMS-02 data \citep[e.g.][]{genolini_bc2,evoli_be, boschini2}.
$D_{xx}$ has a mostly linear relationship at high rigidity.
For our propagation scenario, in all models $\delta$ decreases above a few hundred GV, making the diffusion coefficient steepen, and the propagated CR spectra harden. We confirm that the high-energy break in $\delta$ assumes the recent typical value of $\sim$0.2.  
Figure \ref{fig:dxx} displays the diffusion coefficient as a function of rigidity for A/Z=1 as obtained for our best-fit scenarios. 
For all models, except PD, $D_{xx}$ slightly turns downwards below $\sim1$~GV, as typically expected by standard diffusive-reacceleration scenarios with no wave damping, as \cite[e.g.][and references therein]{cummings_2016, cr_overview}.
The steepening of the turn depends on the best-fit value of $\eta$, which varies across the different models. This parameter only affects the low-energy behavior and is partially degenerate with other parameters that affect low-energy spectra such as the low-energy injection spectral indices. In our models $\eta$ is allowed to assume any value, negative, positive, or null. We find that for all models 0~$<$~$\eta$~$\le$~1. The PD model has a value of $\eta$ very close to 0. This causes its diffusion coefficient to follow a simple power law, not decreasing at low rigidity compared to the other models, but neither increasing as expected by standard plain diffusive models \citep{cr_overview}. In more detail, differently from standard plain diffusion models, we find that present data do not need to assume a break of the diffusion coefficient at low energy in order to fit B/C. Indeed, it has been a commonly accepted long-standing true \cite[e.g.][and references therein]{cr_overview,grenier_strong_2015, strong_2010} that standard plain diffusion models require a break in the diffusion coefficient that produces a strong upwards turn in $D_{xx}$ below a rigidity of a few GV. This has also been  recently confirmed by \citep{cummings_2016} with Voyager and Pamela data and by \citep{genolini_bc2, weinrich_2020} with AMS-02 data. Our PD model does not require such an upturn, so it is closer to our other models than to standard plain diffusion models with the break. This is a very new result that puts into question commonly accepted truths (for a complete discussion see also Section \ref{sec:low_energy_break}). Figure \ref{fig:dxxAll} shows the diffusion coefficient as a function of rigidity for our models compared with the works mentioned above. PD models are from \cite[][PD1]{cummings_2016}, \cite[][PD]{ptuskin_damping}, \cite[][SLIM]{genolini_bc2} and \cite[][SLIM]{weinrich_2020}. It is clear that previous PD models required an upturn or the diffusion coefficient at low energies.

\subsection{Low-energy breaks in the injection spectra}\label{sec:low_energy_break}
Before Voyager CR data became available, many works in the literature \citep[e.g][]{strong_2010,drury_strong_2015}
have shown that B/C and other primary-to-secondary ratios could be reproduced with a diffusive-reacceleration scenario, with no ad hoc low-energy break in the diffusion coefficient. However, they needed a low-energy CR injection break. 
It was also accepted \citep[see e.g.][and references therein]{strong_2010,Vladimirov_2012,grenier_strong_2015,gabici_2019}
that plain diffusion models require a low-energy break in the diffusion coefficient or, alternatively, an additional low-energy CR injection break with respect to diffusive-reacceleration models. 
The break in the diffusion coefficient at a few GV was required to fit the observed decrease of B/C at low energies, which is faster than the $\beta$ dependence.
Recent studies using Voyager data \citep[e.g.][and references therein]{cummings_2016,boschini2} 
have shown that diffusive-reacceleration models require two breaks at low energy, while plain diffusion models \citep[e.g.][]{cummings_2016} 
need also the break in the diffusion coefficient in addition to the two breaks in the injection spectrum. PD and DRE scenarios better fit Voyager data with two breaks in the low-energy injection spectra.

Because of the nonphysical nature of such breaks, in the literature there are many attempts to find a more physical explanation, such as wave damping or local sources. Indeed, even though mechanisms for a low-energy break have been proposed \citep{ptuskin_damping},
the details are still unclear, and a recent work has questioned the necessity of such breaks in explaining Voyager data \citep{phan_voy}, one of which can be possibly attributed to the discrete nature of CR sources.

In this work we confirm that two breaks in the low-energy injection spectra improve the fit of AMS-02 and Voyager data for all models, but at the cost of adding complexity and at the risk of a non convincing physical explanation. As shown for the DRC1 model for illustration, we find that, although being disfavored, one break may still produce an acceptable fit to data for all scenarios, especially considering uncertainties given by the force-field approximation for the modulation. 

As an additional novelty, for plain diffusion models we find that not only is a break in the diffusion coefficient at low energy not needed, but also the PD model needs the same number of low-energy breaks in the injection spectrum as the DRC and DRE scenarios.

\begin{table*}
\centering
\renewcommand{\thefootnote}{\arabic{footnote}}
\caption{Injection spectra for all models.}
%\begin{ruledtabular}
\begin{tabular}{ccccccc}
\hline
Parameter&DRC1&DRC2&DRC\_conv&DRE&PD&DC\\
\hline

$\gamma_{0, H}$ & 1.9498 & 2.0651 & 2.2192 & 2.0750 & 1.8651 & 1.8177\\
$\gamma_{1, H}$ & 2.4445 & 1.9103 & 1.6503 & 1.9050 & 1.9475 & 1.9398\\
$\gamma_{2, H}$ & ˙˙˙    & 2.4165 & 2.4595 & 2.4148 & 2.4460 & 2.4510\\
$\rho_{1, H}\ (GV)$  & 9.7724 & 1.8728 & 0.950 & 0.9512 & 0.950 & 0.950\\
$\rho_{2, H}\ (GV)$  & ˙˙˙ & 10.8912 & 6.970 & 9.3386 & 6.970 & 6.970\\
\hline
$\gamma_{0, He}$ & 1.8425 & 1.8504 & 1.8242 & 1.8344 & 1.5994 & 1.5750\\
$\gamma_{1, He}$ & 2.3763 & 1.9288 & 1.8554 & 1.9691 & 1.9656 & 1.9538\\
$\gamma_{2, He}$ & ˙˙˙    & 2.3495 & 2.3998 & 2.3497 & 2.3767 & 2.3824\\
$\rho_{1, He}\ (GV)$  & 6.9863 & 1.000 & 1.000 & 1.0016 & 0.9854 & 1.000\\
$\rho_{2, He}\ (GV)$  & ˙˙˙ & 8.4194 & 7.490 & 10.2636 & 6.5962 & 7.490\\
\hline
$\gamma_{0, C}$ & 0.8172 & 0.7286 & 1.9321 & 0.9210 & 0.0862 & -0.5963\\
$\gamma_{1, C}$ & 2.3570 & 2.2043 & 0.5733 & 2.2947 & 1.9944 & 2.0575\\
$\gamma_{2, C}$ & ˙˙˙ & 2.3640 & 2.3726 & 2.3601 & 2.3159 & 2.3114\\
$\rho_{1, C}\ (GV)$  & 1.5711 & 1.2670 & 4.7874 & 1.5347 & 0.6510 & 0.7037\\
$\rho_{2, C}\ (GV)$  & ˙˙˙ & 7.9872 & 6.0414 & 12.141 & 8.0493 & 8.7891\\
\hline
$\gamma_{0, O}$ & 1.7627 & 0.9774 & 0.7446 & 0.9855 & 0.9651 & 1.2018\\
$\gamma_{1, O}$ & 2.4750 & 2.0910 & 1.9208 & 2.0186 & 1.8383 & 1.7455\\
$\gamma_{2, O}$ & ˙˙˙    & 2.4267 & 2.4826 & 2.4226 & 2.4645 & 2.5226\\
$\rho_{1, O}\ (GV)$  & 7.6281 & 1.0152 & 0.900 & 0.8993 & 0.8075 & 0.900\\
$\rho_{2, O}\ (GV)$  & ˙˙˙ & 15.4701 & 7.500 & 12.0692 & 7.2230 & 7.500\\
\hline
$\gamma_{0, Ne}$ & 1.6389 & 0.4676 & 0.4827 & 1.5132 & 0.8817 & 0.9941\\
$\gamma_{1, Ne}$ & 2.4653 & 1.6144 & 1.7978 & 1.4924 & 1.6881 & 1.5911\\
$\gamma_{2, Ne}$ & ˙˙˙    & 2.3965 & 2.4766 & 2.3988 & 2.4461 & 2.4544\\
$\rho_{1, Ne}\ (GV)$  & 13.658 & 1.0298 & 1.150 & 0.0737 & 1.150 & 1.150\\
$\rho_{2, Ne}\ (GV)$  & ˙˙˙ & 6.5888 & 9.420 & 6.7613 & 9.420 & 9.420\\
\hline
$\gamma_{0, Mg}$ & 1.1539 & 0.3065 & -0.2246 & 0.9689 & 0.5755 & 0.4308\\
$\gamma_{1, Mg}$ & 2.4827 & 1.8642 & 1.8090 & 1.9735 & 1.8285 & 1.7244\\
$\gamma_{2, Mg}$ & ˙˙˙    & 2.4593 & 2.5346 & 2.4622 & 2.4976 & 2.5042\\
$\rho_{1, Mg}\ (GV)$  & 3.1574 & 1.0070 & 0.850 & 1.4586 & 1.0150 & 0.850\\
$\rho_{2, Mg}\ (GV)$  & ˙˙˙ & 6.8390 & 7.000 & 7.3757 & 7.1691 & 7.000\\
\hline
$\gamma_{0, Si}$ & 1.6127 & 1.2716 & -0.7424 & 0.7955 & 0.8270 & 0.8477\\
$\gamma_{1, Si}$ & 2.4743 & 1.9511 & 1.9095 & 1.9298 & 1.8589 & 1.7883\\
$\gamma_{2, Si}$ & ˙˙˙    & 2.4613 & 2.4700 & 2.4636 & 2.4736 & 2.4719\\
$\rho_{1, Si}\ (GV)$  & 5.3084 & 1.4747 & 0.850 & 1.0043 & 0.9300 & 0.850\\
$\rho_{2, Si}\ (GV)$  & ˙˙˙ & 7.9584 & 7.000 & 9.0610 & 7.6319 & 7.000\\
\hline
\label{table:injection_spectra}
\end{tabular}
%\end{ruledtabular}
\end{table*}

\subsection{Higher-Energy Injection Spectra}

For each model scenario, the injection spectral indices for each primary species are left free in order to have the closest agreement with CR data and to test how the injection spectral indices vary between the different species. In this section we compare the propagated spectral indexes of various species as observed by AMS-02 \citep{ams02_review} 
and our modeled injection spectral indexes that best-fit AMS-02 observations after propagation as reported in Table \ref{table:injection_spectra}. Our aim is to speculate about possible different classes of the primary CR spectra with different rigidity
dependence.

First, the propagated spectrum of He as observed by AMS-02 is harder than the H spectrum above a few GV for the entire energy range covered by AMS-02. We observe the same trend in  our best-fit scenarios: they all have the injection spectrum of He harder than that of H by a factor of about 0.06 for the same energy range, as also found by \citet{evoli_model}.

Second, the AMS-02 collaboration \citep{ams02_review} 
observes that He, C, and O have the same rigidity dependence above 60 GV, and, in addition, that He and C have the same spectral index above 7 GV, so they speculate about their possible common origin.
We confirm the possible same origin of He and C.
Indeed, or all best-fit scenarios, we do observe the trend of He and C forming a group above $\sim7$ GV, while O has a different injection spectral index closer to heavier primary nuclei. 
Regarding O, we point out that similar observed spectral indexes do not mean that the injection spectral indexes are the same, because the contribution from spallation of heavier elements may change the observed propagated spectrum of nuclei, obscuring the possible common origin of various CR species. For example,
He, C, O, Ne, Mg, and Si are usually assumed to be primary nuclei. 
While O satisfies this definition best because the contribution to O from secondaries is expected to be smaller than the data's uncertainties, for C below $\sim$200 GV the contribution to C from secondaries is significant, mainly from fragmentation of O nuclei, and it can even be 20$\%$ - 30$\%$ \citep{evoli_model, tomassetti_2012}.
Hence, if the observed C spectrum is the same as the spectrum of O, this means that the expected injection spectral index of C should be harder than the injection spectral index of O due to the contribution of secondaries at lower rigidity. Hints of this trend are found in our study above 60 GV for all the best-fit scenarios: the injection spectral index of C is harder than that of O 
and it is 
similar to He as found by \citet{ams02_review}.

Third, AMS-02 data suggests that the observed spectral indices are the same for for Ne and Mg above 3.65 GV, and for Ne, Mg, and Si above 200 GV \citep{ams02_review}. 
For our models, 
we do observe that the injection spectral indices of Ne, Mg, and Si are close, and we especially observe a broader grouping including H, O, Ne, Mg, and Si.
Similar results to ours are also found by  \citet{evoli_model}, where they find that adopting three different injection spectra or H, He, and heavier nuclei fits AMS data the best.
Then, they also conclude that, while the injection spectrum varies for H and He, the same spectral index at injection for all heavier nuclei strongly worsens the fit, but it is still statistically accepted by AMS-02 data once the following two conditions apply: uncertainties in the cross sections of production of secondary nuclei are taken into account and the grammage traversed by CRs during transport through the Galaxy is fine-tuned. 
Differently from \citet{evoli_model}, we fit low-energy Voyager data as well, which contribute to better constraining propagation parameters and CR spectral indexes.

\subsection{Positrons and Antiprotons}

Positrons and antiprotons are created by interactions of primary CRs with the interstellar gas rather than being accelerated by shock waves.
Hence, because positrons and antiprotons are supposed to have secondary origin only (except for the positron excess at high energies),
the propagation model parameters are not fit to positron and antiproton spectra. In this section, we show the differences in the modeled positron and antiproton spectra produced by the various propagation scenarios. 

\begin{figure*}
    \centering
    \includegraphics[width=.7\textwidth]{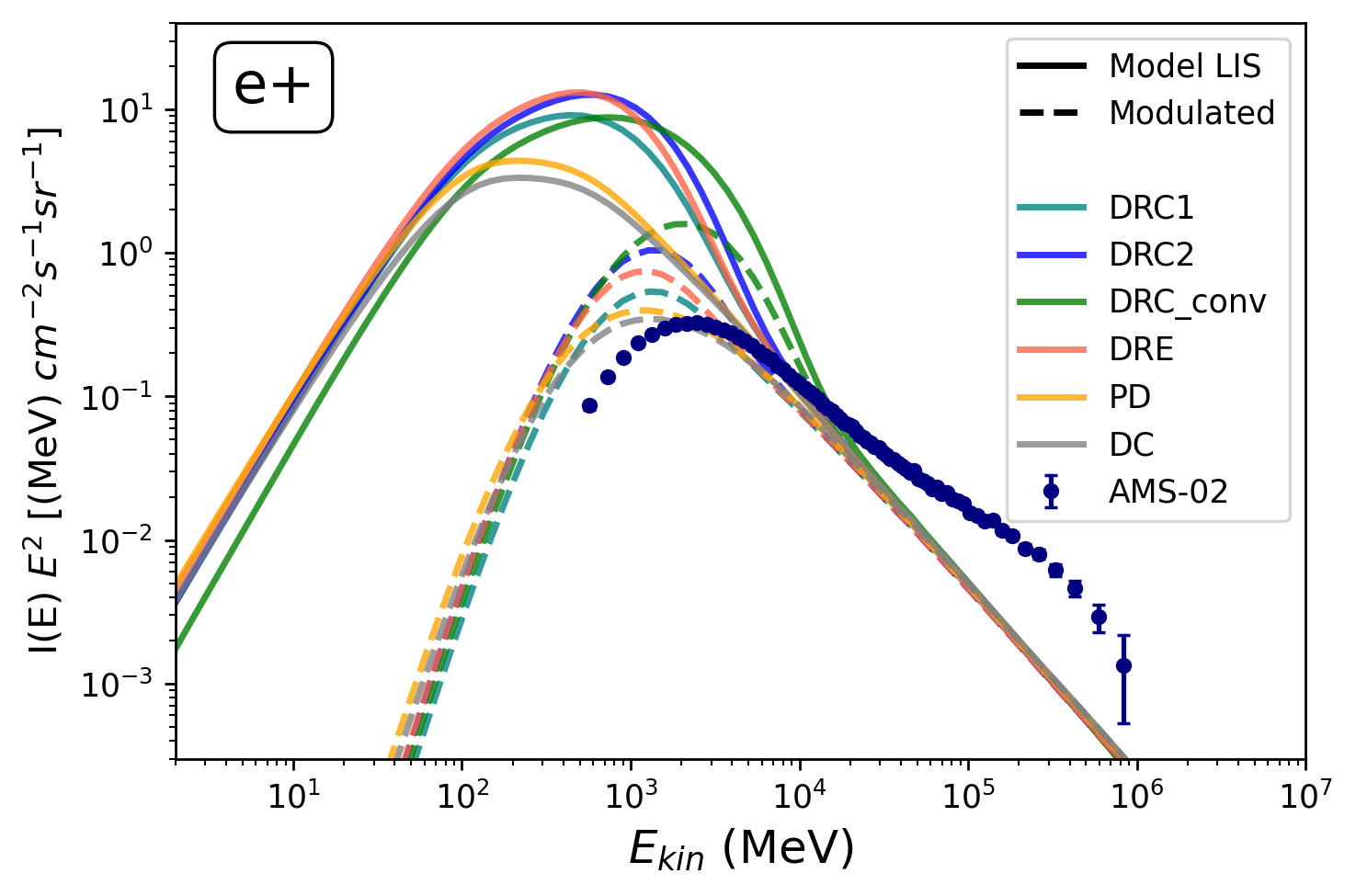}
    \caption{Plot of the e+ spectrum for the six tested scenarios, compared with AMS-02 data \citep{ams02_review}. Solid lines represent the LIS, while dashed lines represent the modulated spectra.}
    \label{fig:e+}
\end{figure*}

Positron spectra compared to AMS-02 positron 
data are shown in Figure \ref{fig:e+} for the LIS (solid lines) and the modulated spectrum (dashed lines) for all scenarios considered. The AMS-02 data plotted is from \citep{ams02_review}. As expected, the models do not perform very well in replicating the data at high energies, when the models fall far below the measured fluxes from AMS-02. This phenomenon is well-observed, and consistent with models that only produce secondary positrons from CR collisions \citep{ams02_review}. The high-energy rise of the positron spectrum \citep{pamela_e+,ams02_e+} is not explained by these models and it is not our purpose here, because it most likely requires new theoretical approaches on the transport of CRs or nearby sources of primaries (for a deeper discussion see e.g. \citet{lipari_article, cr_challenges}). Hence, we confirm that the positron excess cannot be reproduced by any of the standard transport scenarios. 
At low energies, models with reacceleration (DRC and DRE) lead to the greatest predicted positron production. 
In particular, DRC\_conv has a dramatic production of positrons at $\sim1$ GeV, because it is the model with the strongest reacceleration, with $V_{Alf}>50$ km/s. This high production of positrons can lead to an excess of gamma rays and synchrotron emission as compared to available data, which might challenge diffusive-reacceleration models \citep{orlando_cr}. Instead, PD and DC models produce about 3 times lower positrons and a harder spectrum in the 0.1-10 GeV range as also pointed out in \citet{strong_orlando_jaffe_2011, orlando_strong_halo, orlando_cr, orlando_prd}. This would be in agreement with radio, microwave, and gamma-ray data \citep{orlando_cr}. 
Because the modulated spectra are plotted with the same modulation potential value as found for nuclei, the plot clearly shows that the modulation should be stronger for positrons than for hadronic nuclei, especially for the DRC and DRE scenarios. In fact, scenarios with reacceleration would require a much larger solar modulation than PD and DC scenarios.

\begin{figure*}
    \centering
    \includegraphics[width=.6\textwidth]{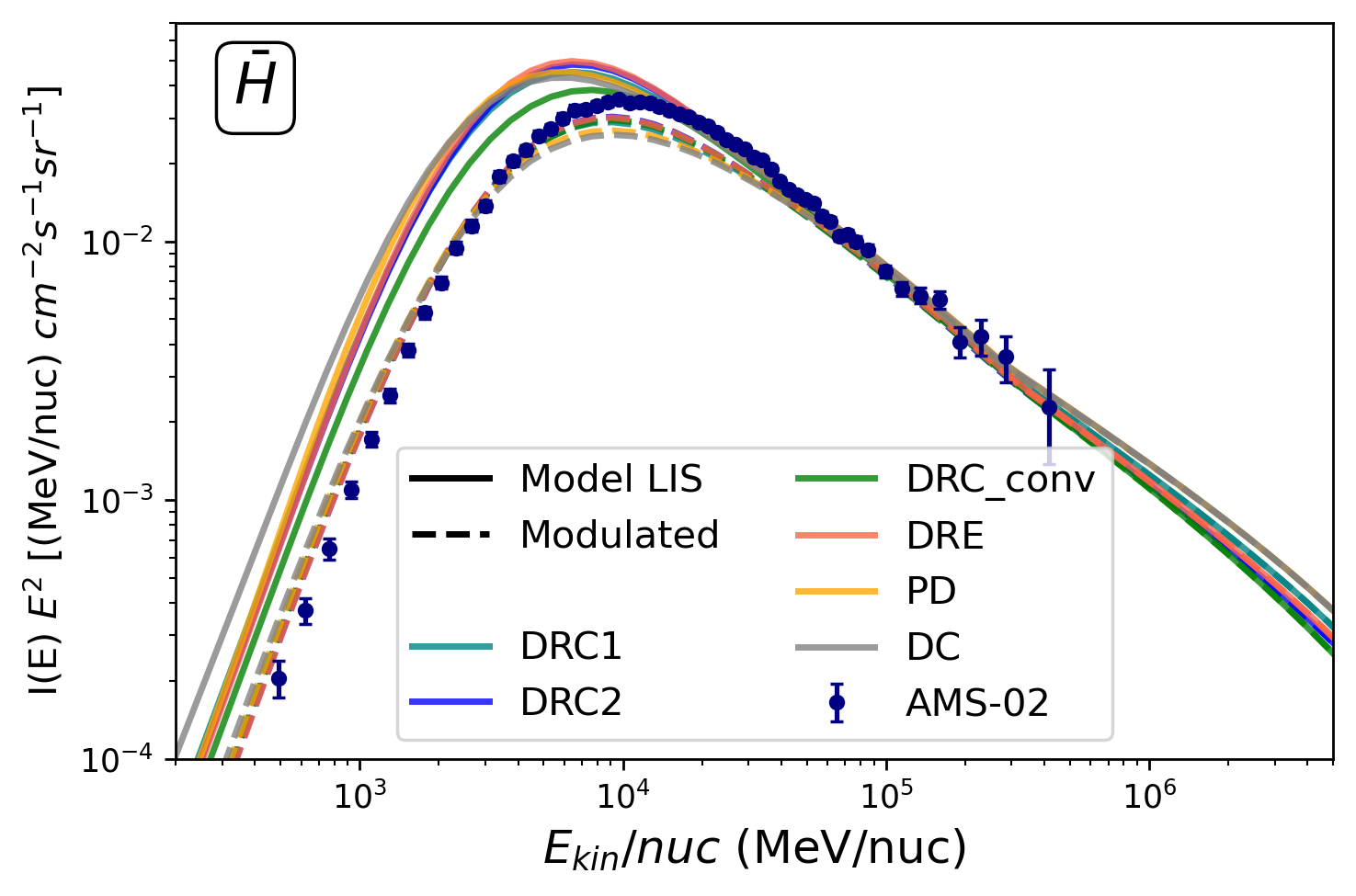}
    \includegraphics[width=.6\textwidth]{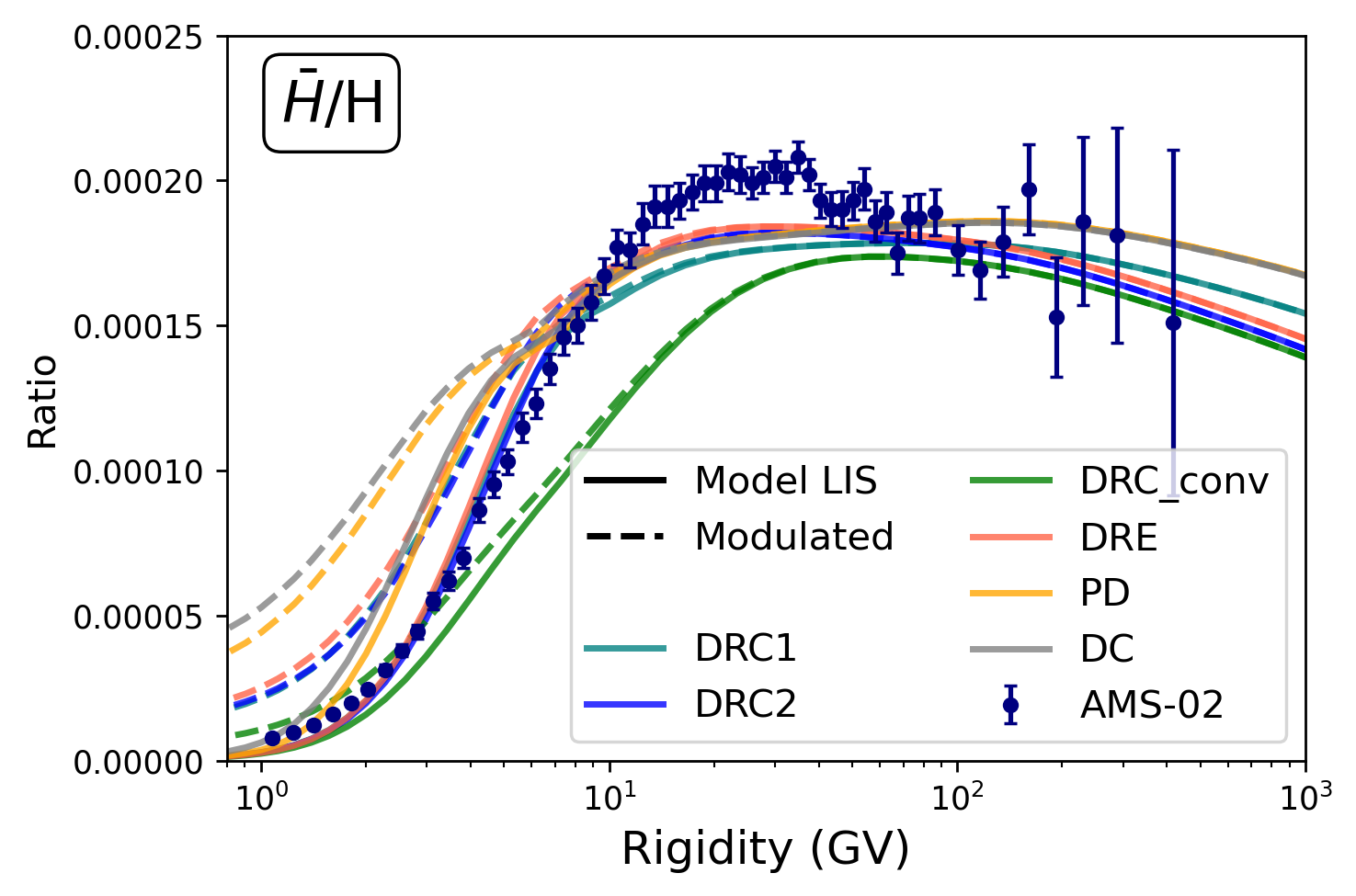}
    \includegraphics[width=.6\textwidth]{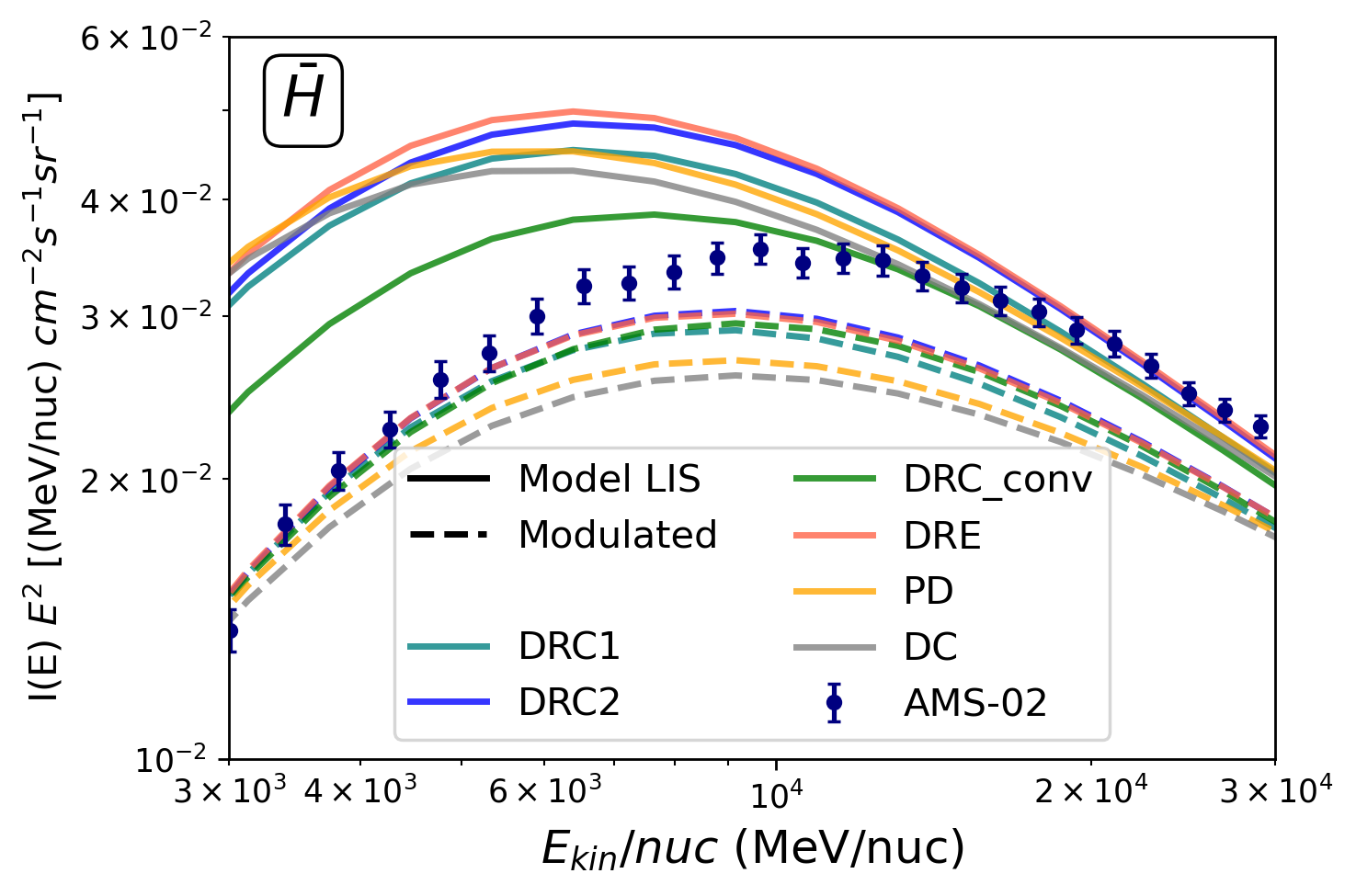}
    \caption{Plot of the $\overline{H}$ spectrum and the ratio of antiprotons to protons for the six tested scenarios, compared with AMS-02 data \citep{ams02_review}. Solid lines represent the LIS, while dashed lines represent the modulated spectra.}
    \label{fig:hbar}
\end{figure*}

Antiproton models compared to AMS-02 antiproton data are shown in Figure \ref{fig:hbar}, together with the ratio of antiprotons to protons. %Here, the modulated spectra have been obtained by fitting the modulation potential independently from the modulation potential of nuclei.
The AMS-02 data plotted is from \citet{ams02_review}.
Antiprotons are calculated by using the cross sections as in \citet{kachelriess_2011,kachelriess_2015}.
At high energies our models of antiprotons have a close agreement with data within the error bars. The modeled antiproton-to-proton ratio has nearly constant rigidity dependence in the range $\sim$40 GV to $\sim$400 GV as found in data \citep{ams02_pbar}. Contrary to some previous studies, there is no need of a new source to explain the high-energy ratio, especially for the PD scenario which does not show clear rigidity dependence in this range. Future and more precise measurements of antiprotons and antiproton-to-proton ratio above several tens of GeV will help to differentiate among the different scenarios.

To account for possible charge dependence modulation the modulated spectra of antiprotons have been obtained by fitting the modulation potential independently from the modulation potential of nuclei. Modulated spectra are shown as dashed lines.
Models of antiprotons generally do not fit the antiproton flux at low energies, in the regions where solar modulation has a significant effect. In more detail, several groups have noticed an excess in the $\sim$10 GeV region, where the associated systematic uncertainties have impeded a common statement.
A subdominant primary component of antiprotons in the 10~--~20 GV range has  been hypothesized by many authors to originate from dark matter annihilation.
Indeed, we find that the best-fit modulated spectra strongly underestimate the flux at $\sim$ 10 GeV/nuc for all propagation scenarios, producing an excess (see the zoom in this region in Figure \ref{fig:hbar}, bottom).
However, due to uncertainties by the solar modulation we are not able to confirm or discharge a dark matter origin for such an excess. 
The difference between the modeled LIS antiproton spectra produced by the various propagation scenarios is a factor of $\sim$1.4 at several GeV/nuc and $\sim$1.8 at 1~TeV/nuc.

\section{Discussion and Conclusions}

\begin{table*}
\caption{Qualitative comparison of models}
%\begin{ruledtabular}
\begin{tabular}{lll}
\hline
Model & Pros & Cons\\
\hline
DRC1 & Many fewer parameters than DRC2 & Slightly worse B/C performance than DRC2\\
DRC2 & Best performance on B/C and other species & Most parameters\\
DRC\_conv & Decent performance  & Worse than others especially than equivalent DRC2 \\
DRE & Simpler model than DRC, but second-best $\chi^2$ & Slightly worse overall performance than DRC2\\
PD & Simplest model, no low-energy diffusion break & Worse performance than DRE and DRC2, esp. on B,C,O\\
DC & Simpler model than DRC & Worse performance than PD, DRE, DRC1, DRC2\\
\hline
\label{table:pros_cons}
\end{tabular}
%\end{ruledtabular}
\end{table*}

This work presents a comparison of different scenarios of CR propagation, and explores their performance compared with recent CR spectrum data. 
This section lists and discusses the results.
\begin{itemize}
\item[-] All best-fit models can explain measurements sufficiently, but the degree of accuracy of the different models helps to inform which scenarios are favored by the Voyager and AMS-02 CR measurements. 
Table \ref{table:pros_cons} summarizes a qualitative comparison among the fitted models. 
Overall all scenarios are able to reproduce Voyager and AMS-02 data, in general, the highest the number of free parameters, the better the chi-square.

For B/C, because the measured cross sections of secondaries are only known with with a few tens per cent error, and even less precisely for some energies and species \citep{evoli_model}, the agreement with B/C data is satisfactory for all the models.

The models DRC1 and DRC2 both agree with AMS-02 and Voyager data well, but DRC2 gives a small improvement at the peak of B/C, hinting for two low-energy breaks in the injection spectrum %at ~0.8 GV and ~7.3 GV, 
instead of one. %at 1.4 GV. 
DRC2 is the best-performing model overall, but DRC1 is a simplification that is still close in performance, while cutting down on the parameter space significantly by only requiring one break and two spectral indices for each CR species.

Our results also include that the currently used range of values for $dV_\textrm{conv}/dz$ around 6-12 km/s/kpc is favored by the data, with the tested model DRC\_conv (with $dV_\textrm{conv}/dz$ $>$ 50 km/s/kpc) having a worse alignment with data.

The DRE scenario, which does not include convection, also can explain the data quite well, with comparable performance than DRC2. 

The PD scenario, which does not include either convection and reacceleration, can also adequately explain the data, but does somewhat worse than the best-fit DRC2 model.

Both PD and DRE perform well at high energies, but are primarily surpassed at low energies by the DRC2 model, with the highest mumber of free model parameters.

The exploration of the possibility of the DC scenario shows that it is disfavored by the data, and the PD scenario is preferred over the DC. The best-fit DC models are just PD models with $dV_\textrm{conv}/dz$=0, and, hence, at the cost of having one additional  parameter. 
The addition of convection only leads to an improvement in performance alongside reacceleration, in the case of moving to a DRC model.

\item[-] Contrary to what usually assumed, the PD scenario does not need an upturn in the diffusion coefficient at low energy to fit B/C, while it needs the same number of low-energy breaks in the injection spectrum as DRE and DRC scenarios.

\item[-] We find that the force-field approximation describes well the differential intensity spectra for
the entire AMS-02 and Voyager energy band and all the analyzed nuclei, except for antiprotons.

\item[-] Antiprotons show severe underestimation at $\sim$ 10 GeV, with an interpretation that remains unclear. Hence, we confirm the $\sim$10 GeV excess in the antiproton spectra for all scenarios.

\item[-] Above $\sim$40~GV there is no need of a new source to explain the antiproton to proton ratio, especially for the PD scenario which does not show a clear rigidity dependence in this range. Future and more precise measurements of antiprotons and antiproton to proton ratio above several tens of GeV will help to differentiate among the different scenarios.

\item[-] By modeling secondaries particles, such as antiprotons and positrons, we found that different scenarios produce positrons that differ for one order of magnitude at $\sim$1 GeV, while different scenarios produce antiprotons that differ for a factor of 2 at several GV.

\item[-] We confirm that the injection spectrum of He should be harder than that of H in order to fit the data.

\item[-] We also find that He and C should have the same injection spectral index above several GV in order to fit  AMS-02 data.

\item[-] In order to fit AMS-02 data O should have a slightly softer injection spectral index than He and C above several GV possibly due to the contribution of secondaries. 

\item[-] For all propagation scenarios, the resulting modulation should be stronger for positrons than for nuclei, with reacceleration models requiring a much larger modulation

\end{itemize}

In this work we show uncertainties given by various propagation scenarios, especially on the estimate of secondaries such as positrons and antiprotons.  
We exploit the different propagation scenarios, while we neglect uncertainties in the cross sections. 
It has been established that one of the primary next steps to improve our understanding of CR propagation and subtle effects is to reduce uncertainties in the cross sections.  
Other approaches by using different sources of data simultaneously, including current and future direct CR measurements as well as synchrotron radiation and gamma-ray measurements have also been put forward. %\cite{strong_orlando_jaffe_2011,orlando_cr}. 

This work contributes to the exploration of CR propagation models, and further continuing coherent work is necessary to fully refine our understanding of CR and their propagation in the Galaxy.

%%%%%%

\acknowledgments

This work was supported in part by the U.S. Department of Energy, Office of Science, Office of Workforce Development for Teachers and Scientists (WDTS) under the Science Undergraduate Laboratory Internships (SULI) program. This work was supported also by NASA grant No. 80NSSC22K0495.

\bibliography{bibliograph}{}
\bibliographystyle{aasjournal}



\end{document}